\documentclass[aps,reprint,groupedaddress,floatfix,showpacs]{revtex4-1}

\usepackage{graphicx,color}
\usepackage{bm}
\usepackage{amsmath,amssymb}
\usepackage{array}
\usepackage{multirow}
\usepackage{braket}
\newcommand{\mr}[2]{#1 \  \mathrm{#2}}
\newcommand{\ang}{\mbox{\AA}}

\begin{document}

\title{Magnetic hexadecapole order and magnetopiezoelectric metal state in Ba$_{1-x}$K$_x$Mn$_2$As$_2$ } 
\author{Hikaru Watanabe}
\email[]{watanabe.hikaru.43n@st.kyoto-u.ac.jp}
\author{Youichi Yanase}

\affiliation{Department of Physics, Graduate School of Science, Kyoto University, Kyoto 606-8502, Japan}


\begin{abstract}
We study an odd-parity magnetic multipole order in Ba$_{1-x}$K$_x$Mn$_2$As$_2$ and related materials. Although $\rm BaMn_2As_2$ is a seemingly conventional Mott insulator with G-type antiferromagnetic order, we identify the ground state as a magnetic hexadecapole ordered state accompanied by simultaneous time-reversal and space-inversion symmetry breaking. A symmetry argument and microscopic calculations reveal the ferroic ordering of leading magnetic hexadecapole moment and admixed magnetic quadrupole moment. Furthermore, we clarify electromagnetic responses characterizing the magnetic hexadecapole state of semiconducting $\rm BaMn_2 As_2$ and doped metallic systems. A magnetoelectric effect and antiferromagnetic Edelstein effect are shown. Interestingly, a counter-intuitive current-induced nematic order occurs in the metallic state. The electric current along the \textit{z}-axis induces the \textit{xy}-plane nematicity in sharp contrast to the spontaneous nematic order in superconducting Fe-based 122-compounds. Thus, the magnetic hexadecapole state of doped BaMn$_2$As$_2$ is regarded as a \textit{magnetopiezoelectric metal}. 
Other candidate materials for magnetic hexadecapole order are proposed.
\end{abstract}

\maketitle

\section{Introduction}
Multipole moment, a concept established in the classical electromagnetism, characterizes the anisotropy of electric and magnetic charge distribution. Emergent multipole order in condensed matter physics has attracted fundamental interests for more than three decades~\cite{kuramoto2009}. Ferroic and antiferroic order of multipole moment has been observed in many $d$- and $f$-electron systems. Although previous studies have focused on the even-parity multipole order~\cite{kuramoto2009}, recent studies point to the \textit{odd-parity multipole order} which may be realized in locally noncentrosymmetric systems~\cite{hitomi2014,yanase2014,hayami2014,hayami2016,sumita2016,hitomi2016,sumita2017}.
Experimentally, several materials have been identified~\cite{fiebig2005}, which can be traced back to Cr$_2$O$_3$~\cite{astrov1960}.

Locally noncentrosymmetric systems preserve global inversion symmetry in the crystal structure although the local site symmetry lacks inversion symmetry. Then, the antisymmetric spin-orbit coupling (ASOC) entangles various degrees of freedom such as spin, orbital, and sublattice~\cite{Maruyama2012,Fischer2011}. The peculiar electronic structure may cause intriguing phenomena characterizing odd-parity multipole order, such as magnetoelectric (ME) effect~\cite{hayami2014,yanase2014}.
Although previous theoretical works of odd-parity magnetic multipole order are based on toy models~\cite{yanase2014,hayami2014,hayami2016,sumita2016}, in this paper we show the complete classification of magnetic multipole order in tetragonal systems and identify the magnetic hexadecapole order in $\rm BaMn_2As_2$. Characteristic electromagnetic responses in the magnetic hexadecapole state are clarified. 

\begin{figure}[htbp] 
\centering 
\includegraphics[width=70mm,clip]{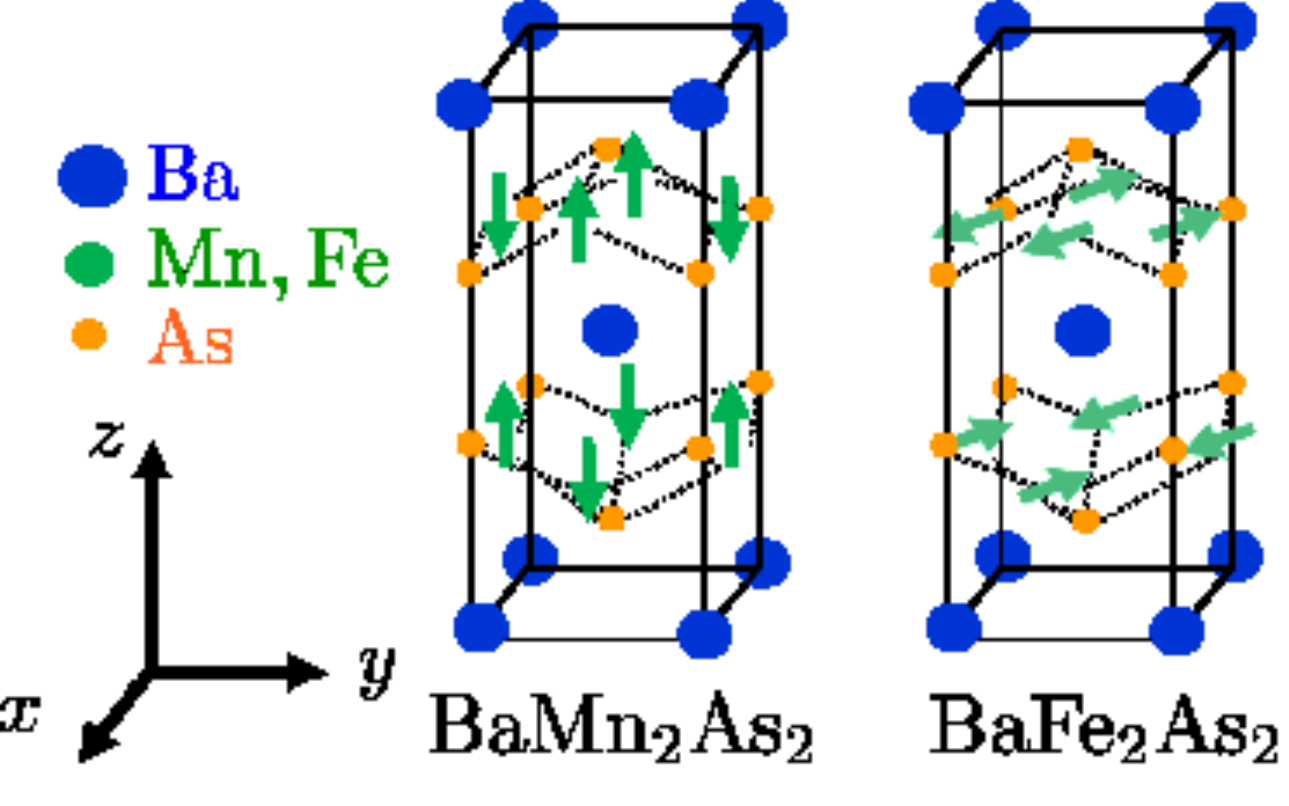} 
\caption{(Color online)
Contrast of magnetic structure of $\rm BaMn_2As_2$ and $\rm BaFe_2As_2$.} 
\label{magcomparison} 
\end{figure}

$ \rm Ba Mn_2As_2$ is an isostructural compound of $\rm Ba Fe_2As_2$, a parent compound of Fe-based high-temperature superconductors (the space group is No.139, $I4/mmm$). However, physical properties of $\rm Ba Mn_2As_2$ and doped $\rm Ba Mn_2As_2$ are quite different from the Fe-based compounds; $ \rm Ba Mn_2As_2$ undergoes the G-type antiferromagnetic (AFM) transition below high N\'eel temperature $T_{\rm N} =\mr{625}{K}$ and  shows semiconducting behaviors~\cite{singh2009a,singh2009b,johnston2010}. On the other hand, $ \rm BaFe_2As_2$ is a metallic compound with a stripe magnetic structure~\cite{johnston2010,ishida2009,dai2015} shown in Figure~\ref{magcomparison}. Neither superconductivity nor structural transition, which have been observed in Fe-based 122 compounds~\cite{johnston2010,ishida2009,dai2015}, occurs in doped $\rm Ba Mn_2As_2$. 

The ground state of $\rm Ba Mn_2As_2$ seems to be a conventional Mott insulator with AFM order~\cite{yao2011,mcnally2015}, analogous to cuprate high-temperature superconductors. However, we notice unusual symmetry of the AFM state, namely, unbroken translation symmetry. This is, indeed, because of a locally noncentrosymmetric crystal structure of $\rm BaMn_2 As_2$. The two Mn sites are crystallographically nonequivalent even in the paramagnetic state. In the folded Brillouin zone, the wave vector of magnetic order is $\bm{q}=0$, and therefore, a ferroic order parameter may characterize the seemingly "AFM order". Because the space-inversion (SI) symmetry is broken instead of the translation symmetry, an odd-parity multipole moment may be a relevant order parameter specifying the ground state of BaMn$_2$As$_2$. Interestingly, $\rm BaMn_2 As_2$ can be metalized by doping hole carriers (Ba$_{1-x}$\textit{A}$_{x}$Mn$_2$As$_2$, \textit{A}= K, Rb) \cite{pandey2012,bao2012,yeninas2013,pandey2013,pandey2015} or applying high pressure~\cite{satya2011}. Then, the AFM order is robust in the  hole-doped regime~\cite{pandey2012,yeninas2013,pandey2013,pandey2015,lamsal2013}. Hence, unconventional responses characteristic of itinerant odd-parity magnetic multipole state are expected, for which studies may open a new paradigm of multipole physics.
 
 The paper is organized as follows. In Sec.~II, we classify the magnetic multipole order by group theory and identify the candidates of order parameter in $\rm BaMn_2As_2$. A complete classification based on irreducible representations (IRs) of a given point group is carried out as done for unconventional superconductors~\cite{Sigrist-Ueda}.  In Sec.~III, we microscopically evaluate the magnetic multipole moment. There remains an ambiguity of the definition for odd-parity magnetic multipole moment in crystals, similar to electric polarization in a bulk system. To avoid this difficulty, we propose a unique definition of odd-parity magnetic multipole moment by difference from a reference state. In Sec.~IV, we introduce an effective single-band model Hamiltonian for studies of electromagnetic responses. In Sec.~V, we demonstrate ME effect arising from the magnetic hexadecapole order and its enhancement in the metallic state. The AFM Edelstein effect is also shown. In Sec.~VI, we show a counter-intuitive current-induced nematic order, the in-plane (\textit{xy}-plane) rotational symmetry breaking by out-of-plane electric current (${\bm J} \parallel \hat{z}$). This response is a manifestation of odd-parity magnetic order in the metallic system. In Sec.~VII, a brief summary is given, and we propose other magnetic hexadecapole compounds showing magnetic structure similar to BaMn$_2$As$_2$.

\section{Group-theoretical classification}
In general, a phase transition leads to symmetry reduction, such as the time-reversal (TR) symmetry breaking by ferromagnetic order. The crystal symmetry of the system is represented by point group, and thus phase transitions can be characterized by the reduction of the point group. In the framework of the group theory, physical quantities are classified into IRs of a given point group, and symmetry constraints for emergent responses are obtained. The order parameter of the phase transition has to belong to the totally-symmetric IR of the point group in the ordered state, but not in the normal state. This scheme is supported by Landau's symmetry argument of second-order phase transitions~\cite{landau1937}.

In $\rm BaMn_2 As_2$, the crystallographic point group $D_{4h}$ descends to the sub-group $D_{2d}$ by the AFM transition. IRs in the normal state are reduced to those in the ordered state as shown in Table \ref{d4h_d2d}. The IRs of $D_{2d}$ do not have subscripts $g / u$,  which indicate the SI symmetry breaking by the AFM order. Hence, it is suggested that the seemingly conventional G-type AFM order of $\rm BaMn_2As_2$ is identified as a parity-violating odd-parity multipole order. According to Table \ref{d4h_d2d}, only the $B_{1u}$ IR in the normal state is reduced to the totally-symmetric $A_1$ IR in the AFM state. Following the group-theoretical framework, we conclude that a basis function belonging to the $B_{1u}$ IR is a relevant order parameter of BaMn$_2$As$_2$.

\begin{table}[htbp]
 \caption{Reductions of IRs $ D_{4h} \rightarrow D_{2d}$. The two-fold rotational symmetry axes of $D_{2d}$ are the $x$/$y$ axes of $D_{4h}$.    }
\label{d4h_d2d}
\begin{tabular}{c|cccccccccc} 
 $D_{4h}$&$  A_{1g}  $&$  A_{2g}  $&$  B_{1g}  $&$ B _{2g}  $&$ E _{g}  $&$  A_{1u}  $&$  A_{2u}  $&$   B_{1u} 	  $&$ B _{2u}  $&$ E _{u}  $ \\ \hline
 $D_{4h} \downarrow D_{2d}$&$  A_{1}  $&$  A_{2}  $&$  B_{1}  $&$ B _{2}  $&$ E   $&$  B_{1}  $&$  B_{2}  $&$   A_{1}   $&$ A _{2}  $&$ E   $
\end{tabular}
 \end{table}
 
Now, we classify magnetic multipole moments in the tetragonal system with the $D_{4h}$ symmetry, and make a list of possible magnetic multipole order. 
 The magnetic multipole moments are written as \cite{kusunose2008} 
	\begin{equation}
	M_{lm} =\mu_{\rm B} \sum_{i} \left(   \frac{2 \bm{l}^{(i)}}{l+1} +2 \bm{s}^{(i)}     \right)\cdot \nabla_i \left( 	r_i^l  \sqrt{\frac{4\pi}{2l+1}  } Y_{lm}^{\ast} (\theta_i ,\phi_i ) 	\right), \label{magmulti}
	\end{equation}
where $\mu_{\rm B}$, $\bm{l}$, $\bm{s}$, and $Y_{lm}$ are respectively Bohr magneton, orbital angular momentum, spin, and spherical harmonics. The label $i$ represents electrons in the unit cell and $( r_i, \theta_i, \phi_i )$ are polar coordinates of the $i$-th electron from a reference point. The phase factor satisfies $Y_{lm}^\ast = (-1)^l Y_{l-m}$ (Condon-Shotley phase). When we discuss multipole moments in a lattice system, it is convenient to use cubic harmonics $Z_{lm}^{\pm}$ defined by
	\begin{equation}
	\begin{split}
		Z_{lm}^+ = \frac{(-1)^m}{\sqrt{2}} \left( Y_{lm} + Y_{lm}^\ast \right), \\
		Z_{lm}^{-} = \frac{(-1)^m}{i\sqrt{2}} \left( Y_{lm} - Y_{lm}^\ast \right),
	\end{split}\label{cubic}
	\end{equation}
for $0< l$ and $0 < m \leq l$. When $m=0$, we denote $Z_{l0}=Y_{l0}$. Accordingly, the multipole moment in the Cartesian coordinates is denoted by 
	\begin{equation}
	\begin{split}
		M_{lm}^+ = \frac{(-1)^m}{\sqrt{2}} \left( M_{lm} + M_{lm}^\ast \right), \\
		M_{lm}^{-} = \frac{(-1)^m}{i\sqrt{2}} \left( M_{lm} - M_{lm}^\ast \right).
	\end{split}\label{magcubic}
	\end{equation}
In our classification, we treat the spin and orbital angular momentum as a classical axial-vector, 
$\left( \hat{\bm{x}}, \hat{\bm{y}}, \hat{\bm{z}} \right) \equiv \mu_{\rm B} \left( 2 \bm{l}/(l+1) +2\bm{s} \right)$, since we take thermodynamical and quantum mechanical expectation values. 
Table \ref{d4h_magmulti} shows the classification of multipole moments of low rank ($l\leq 4$) in the $D_{4h}$ point group symmetry, revealing candidates of order parameter of the AFM state in $\rm BaMn_2As_2$. Up to rank-4, magnetic multipole moments belonging to the $B_{1u}$ IR of the $D_{4h}$ point group are
	\begin{align}
	&M_{22}^+ \  \text{(Quadrupole)}:\sqrt{3}  \left( 	 x\hat{\bm{x}}-  y\hat{\bm{y}} 	\right),	 \label{quad}\\
	 &M_{42}^+ \ \text{(Hexadecapole)}:
	 \begin{cases}
	3 \sqrt{5}   z  \left(	x^2-y^2	\right) \hat{\bm{z}} \\
	 + \frac{\sqrt{5}}{2} \left(7z^2- r^2\right) \left( x \hat{\bm{x}}- y \hat{\bm{y}} 	\right) \\
	  -\frac{ \sqrt{5}}{2}   \left(x^2 -y^2 \right)   \left( x\hat{\bm{x}} +y\hat{\bm{y}} \right).
	\end{cases} \label{hexadeca}
	\end{align}
These basis functions are certainly invariant under all symmetry operations in the  AFM state. Thus, the AFM order may be identified as magnetic quadrupole order or magnetic hexadecapole order. In Sec.~III, we microscopically evaluate multipole moments and show that the magnetic hexadecapole moment is the leading order parameter.
 
In Table \ref{magorder}, we show the complete classification of magnetic multipole order parameter in the $D_{4h}$ point group. The TR odd basis functions in both real space and momentum space are listed. In the real space representation, the basis functions are nothing but the magnetic multipole moments. On the other hand, the momentum space representation looks quite different from the real space representation for the odd-parity magnetic multipole order. This is because the parity under TR operation is opposite between $\bm{r}$ and $\bm{k}$. It should be noticed that the odd-parity basis functions in the momentum space are "spin-independent". They indicate spin-independent corrections to the energy spectrum, which are characteristic feature of odd-parity magnetic multipole states. Although both TR and SI symmetries are broken, the combined {\it PT} symmetry is preserved. Therefore, the Kramers degeneracy at each momentum is ensured, and the deformation of band structure has to be spin-independent.


\onecolumngrid

		\begin{table}[htbp]
		\caption{List of magnetic multipoles up to rank-4. First and second column show a rank and symbol of magnetic multipole, respectively. Third column shows IR in the point group $D_{4h}$. Fourth column shows a representation by local magnetic moment [Eq.~\eqref{magmulti}]. We also show toroidal dipole moment $T_i$ and magnetic monopole moment $\bm{r}\cdot\bm{s}$,  which are not represented by any linear combination of magnetic multipole moment.}
 		 \label{d4h_magmulti}
		 {\renewcommand \arraystretch{1.35}
		 \begin{tabular}{c|c|c|c}
		 $l$&$M_{lm}$&IR& basis function  \\ \hline \hline
		$l=1$&$M_{10}	$&$ A_{2g}$&$\hat{\bm{z}}		$\\
		&$M_{11}^{+}	$&$ E_{g}$&$ \hat{\bm{x}} 		$\\
		&$M_{11}^{-}	$&$ E_{g}$&$ \hat{\bm{y}}		$\\ \hline
		$l=2$&$M_{20}		$&$ A_{1u}$&$   2z\hat{\bm{z}} - x\hat{\bm{x}} 	- y\hat{\bm{y}} $\\
		&$M_{21}^{+}		$&$ E_{u}$&$\sqrt{3}  \left( 	 x\hat{\bm{z}}+  z\hat{\bm{x}} 	\right)		$\\
		&$M_{21}^{-}		$&$ E_{u}$&$\sqrt{3}  \left( 	 y\hat{\bm{z}}+  z\hat{\bm{y}} 	\right)		$\\
		&$M_{22}^{+}		$&$ B_{1u}$&$\sqrt{3}  \left( 	 x\hat{\bm{x}}-   y\hat{\bm{y}} 	\right)		$\\
		&$M_{22}^{-}		$&$ B_{2u}$&$\sqrt{3}  \left( 	  y\hat{\bm{x}}+  x\hat{\bm{y}} 	\right)		$\\ \cline{2-4}
		&$T_x 				$&$  E_{u}	$&$	 z\hat{\bm{y}}- y\hat{\bm{z}}						$\\
		&$T_y 				$&$  E_{u}	$&$	 x\hat{\bm{z}}- z\hat{\bm{x}}						$\\
		&$T_z 				$&$  A_{2u}	$&$	 y\hat{\bm{x}}- x\hat{\bm{y}}						$\\
		&Monopole		&$  A_{1u}	$&$	 x\hat{\bm{x}}+ y\hat{\bm{y}} + z\hat{\bm{z}}						$\\ \hline
		$l=3$&$M_{30}		$&$ A_{2g}$&$\frac{3}{2}    \left( 3z^2 -r^2 \right) \hat{\bm{z}} -  3z  \left( 	 x \hat{\bm{x}}+  y\hat{\bm{y}} 	\right)		$\\
		&$M_{31}^{+}		$&$ E_{g}$&$2\sqrt{6}  zx\hat{\bm{z}}   + \frac{\sqrt{6}}{4}  \left(5z^2-r^2\right)\hat{\bm{x}} -\frac{\sqrt{6}}{2} x^2   \hat{\bm{x}} -  \frac{\sqrt{6}}{2} xy\hat{\bm{y}} 	$\\
		&$M_{31}^{-}		$&$ E_{g}$&$2\sqrt{6}  yz\hat{\bm{z}}   + \frac{\sqrt{6}}{4}  \left(5z^2-r^2\right) \hat{\bm{y}} -\frac{\sqrt{6}}{2}   y^2 \hat{\bm{y}} -  \frac{\sqrt{6}}{2}  xy \hat{\bm{x}}	$\\
		&$M_{32}^{+}		$&$ B_{2g}$&$\frac{\sqrt{15}}{2} \left(	x^2-y^2	\right)\hat{\bm{z}} +\sqrt{15} z \left(	  x \hat{\bm{x}} - y \hat{\bm{y}}  	\right) 	$\\
		&$M_{32}^{-}		$&$ B_{1g}$&$ \sqrt{15} xy \hat{\bm{z}} +\sqrt{15}   z \left(	 y \hat{\bm{x}}  + x \hat{\bm{y}}	\right)  		$\\
		&$M_{33}^{+}		$&$ E_{g}$&$ \frac{3\sqrt{10}}{4} 	 \left(	x^2-y^2	\right) \hat{\bm{x}}  - \frac{3\sqrt{10}}{2}  xy \hat{\bm{y}}  	$\\
		&$M_{33}^{-}		$&$ E_{g}$&$ \frac{3\sqrt{10}}{4} 	 \left(	x^2-y^2	\right) \hat{\bm{y}}  + \frac{3\sqrt{10}}{2}  xy \hat{\bm{x}}  		$\\ \hline
		%
		$l=4$&$M_{40}		$&$ A_{1u}$&$   2z \left( 5z^2 -3 r^2 \right) \hat{\bm{z}}  -\frac{3}{2}   \left( 	5z^2-r^2 	\right) \left(	x \hat{\bm{x}} + y \hat{\bm{y}} 	\right)  		$\\
		&$M_{41}^{+}		$&$ E_{u}$&$\frac{3\sqrt{10}}{4} 	  x \left(	5z^2-r^2	\right) \hat{\bm{z}}   + \frac{\sqrt{10}}{4}   z \left(7z^2-3 r^2\right) \hat{\bm{x}} -\frac{3\sqrt{10}}{2} zx   \left(  x	 \hat{\bm{x}}   +y \hat{\bm{y}} \right) $\\
		&$M_{41}^{-}		$&$ E_{u}$&$\frac{3\sqrt{10}}{4} 	 y \left(	5z^2-r^2	\right)  \hat{\bm{z}}   + \frac{\sqrt{10}}{4}   z \left(7z^2-3 r^2\right) \hat{\bm{y}} -\frac{3\sqrt{10}}{2} yz   \left(x	\hat{\bm{x}}  +y\hat{\bm{y}}  	\right) $\\
		&$M_{42}^{+}		$&$B_{1u}$&$3 \sqrt{5}   z  \left(	x^2-y^2	\right) \hat{\bm{z}}  + \frac{\sqrt{5}}{2} \left(7z^2- r^2\right) \left( x \hat{\bm{x}}- y \hat{\bm{y}} 	\right)  -\frac{ \sqrt{5}}{2}   \left(x^2 -y^2		\right)  \left(x	\hat{\bm{x}} +y\hat{\bm{y}}  	\right)  $\\
		&$M_{42}^{-}		$&$ B_{2u}$&$6 \sqrt{5}   xyz \hat{\bm{z}}    + \frac{\sqrt{5}}{2} \left(7z^2- r^2\right) \left(	 y \hat{\bm{x}} +  x \hat{\bm{y}}  	\right)  - \sqrt{5}  xy \left(	   x \hat{\bm{x}}+ y \hat{\bm{y}} 	\right)   $\\
		&$M_{43}^{+}		$&$ E_u$&$ \frac{\sqrt{70}}{2}  	   x \left(	x^2- y^2 	\right) \hat{\bm{z}} - \frac{\sqrt{70}}{4}  	   x \left(	x^2+ y^2 	\right) \hat{\bm{z}}  +\frac{3\sqrt{70}}{4}  	   z \left(	x^2-  y^2 	\right) \hat{\bm{x}} -\frac{3\sqrt{70}}{2}  	   xyz \hat{\bm{y}}      $\\
		&$M_{43}^{-}		$&$ E_u $&$ \frac{\sqrt{70}}{2}  	   y \left(	x^2- y^2 	\right) \hat{\bm{z}}  +  \frac{\sqrt{70}}{4}  	   y \left(	x^2+ y^2 	\right) \hat{\bm{z}}   +\frac{3\sqrt{70}}{4}  	   z \left(	x^2-  y^2 	\right) \hat{\bm{y}} +\frac{3\sqrt{70}}{2}  	   xyz \hat{\bm{x}}      $\\
		&$M_{44}^{+}		$&$ A_{1u} $&$  \frac{\sqrt{35}}{4}  \left(	x^2+ y^2 	\right) \left(	   x\hat{\bm{x}} +  y \hat{\bm{y}}	\right) +\frac{\sqrt{35}}{4}  \left(	x^2- y^2 	\right) \left(	   x\hat{\bm{x}} -  y \hat{\bm{y}}	\right)	  - \frac{3\sqrt{35}}{2}  x y \left(	   y\hat{\bm{x}} +  x \hat{\bm{y}}	\right)	  $\\
		&$M_{44}^{-}		$&$ A_{2u} $&$  \frac{\sqrt{35}}{2}  \left(	x^2- y^2 	\right)  \left( y\hat{\bm{x}} +   x \hat{\bm{y}}	\right)	+ \sqrt{35} x y \left(	  x\hat{\bm{x}}  -  y\hat{\bm{y}}	\right)	$\\
		\end{tabular}
	}
		\end{table}

		\begin{table}[htbp]
		\caption{The TR odd basis functions of IRs in $D_{4h}$. Basis are represented both in real space and in momentum space. The totally-symmetric IR ($A_{1g}$) is not shown.}
		\label{magorder}
	 {\renewcommand \arraystretch{1.35}
		\begin{tabular}{c|c|c|c}
		IR &\multicolumn{2}{c|}{Basis in real space}	& Basis in momentum space	 \\ \hline \hline
		\multirow{2}{*}{$A_{2g}$}&$M_{10}  $&$\hat{\bm{z}} $&$\hat{\bm{z}}  	$ \\ 
				&$M_{30}  $&$ z \left(  x\hat{\bm{x}}+y \hat{\bm{y}}	\right), z^2 \hat{\bm{z}} 	$&$k_z \left(k_x \hat{\bm{x}}  +k_y \hat{\bm{y}} \right) , k_z^2 \hat{\bm{z}} $ \\ \hline
		$ B_{1g}$	&$M_{32}^-  $&$xy \hat{\bm{z}} ,z  \left( y \hat{\bm{x}}+x \hat{\bm{y}} 	\right) 	$&$k_x k_y \hat{\bm{z}}   , k_z \left(	k_y \hat{\bm{x}}  +k_y \hat{\bm{y}} \right) 	$ \\ \hline
		$ B_{2g}$ &$M_{32}^+ $&$ \left(	x^2-y^2	\right) \hat{\bm{z}} , z \left(	 x\hat{\bm{x}}- y\hat{\bm{y}}	\right) 	$&$\left(	k_x^2-k_y^2	\right) \hat{\bm{z}}  ,k_z  \left(	k_x \hat{\bm{x}}  -k_y \hat{\bm{y}}   	\right) 	$ \\ \hline
		\multirow{4}{*}{$ E_{g}$}	&$M_{11}^{\pm}   $&$\left[ \hat{\bm{x}}, \hat{\bm{y}} 		\right]  	$&$\left[ \hat{\bm{x}}, \hat{\bm{y}} 		\right]  	$ \\
		&\multirow{1}{*}{$ M_{31}^{\pm}$}&$\left[ zx \hat{\bm{z}} ,  yz \hat{\bm{z}}   	\right],  \left[ z^2 \hat{\bm{x}} , z^2 \hat{\bm{y}}  \right]  $&$\left[ k_zk_x \hat{\bm{z}} ,  k_yk_z \hat{\bm{z}}   	\right],  \left[ k_z^2 \hat{\bm{x}} , k_z^2 \hat{\bm{y}}  \right] $  \\  
		&&$\left[ x^2 \hat{\bm{x}} ,  y^2 \hat{\bm{y}}   	\right],  \left[ xy \hat{\bm{x}} , xy \hat{\bm{y}}  \right]  $&$    \left[ k_x^2 \hat{\bm{x}} ,  k_y^2 \hat{\bm{y}}   	\right],  \left[ k_xk_y \hat{\bm{x}} , k_xk_y \hat{\bm{y}}  \right]$  \\  
					&$M_{33}^{\pm} $&$\left[ xy \hat{\bm{x}} ,  xy \hat{\bm{y}}   		\right] ,\left[ \left(	x^2-y^2	\right) \hat{\bm{x}} ,\left(	x^2-y^2	\right) \hat{\bm{y}}  \right]  $&$  \left[ k_xk_y \hat{\bm{x}} ,  k_xk_y \hat{\bm{y}}   		\right] , \left[ \left(	k_x^2-k_y^2	\right) \hat{\bm{x}} ,\left(	k_x^2-k_y^2	\right) \hat{\bm{y}}  \right] $  \\  \hline
		\multirow{5}{*}{$ A_{1u}$}		&$M_{20} $&$ 2z \hat{\bm{z}}-x \hat{\bm{x}}- y \hat{\bm{y}}	  	$&\multirow{5}{*}{$k_xk_yk_z \left(	k_x^2-k_y^2	\right) $}	 \\
		&Monopole&$x\hat{\bm{x}}+y\hat{\bm{y}}+z\hat{\bm{z}}$&\\
		&$M_{40} $&$  z^3  \hat{\bm{z }},  z^2 \left(	x\hat{\bm{x}} +y \hat{\bm{y}}	\right)			$&\\ 
		&\multirow{1}{*}{$M_{44}^+ $}&$  \left(x^2+y^2	\right)  \left(	 x \hat{\bm{x}} +  y \hat{\bm{y}} \right)	  ,\left( x^2-y^2		\right) \left(	x\hat{\bm{x}} -y \hat{\bm{y}}  \right)		$&\\ 
		&&$  xy \left(	y\hat{\bm{x}} +x \hat{\bm{y}}  \right)		$&\\ \hline
		\multirow{2}{*}{$ A_{2u}$}			&$T_z$&$	 y\hat{\bm{x}}- x\hat{\bm{y}}	$&\multirow{2}{*}{$k_z $} \\ 
		&$M_{44}^- $&$ \left(x^2-y^2	\right)  \left(	 y \hat{\bm{x}} +  x \hat{\bm{y}} \right) ,	xy \left(	x\hat{\bm{x}}  - y\hat{\bm{y}} \right)	$&\\ \hline
		\multirow{3}{*}{$ B_{1u}$}	&$M_{22}^+ $&$    x\hat{\bm{x}} -y \hat{\bm{y}} 	$& \multirow{3}{*}{$k_xk_yk_z    $} \\ 
					&\multirow{1}{*}{$ M_{42}^+$}&$ 	 z \left(	x^2-y^2	\right)  \hat{\bm{z}}  	$& \\  
					&&$ 	z^2 \left(	x \hat{\bm{x}}- y \hat{\bm{y}}  \right), \left(x^2-y^2	\right) \left(	 x  \hat{\bm{x}} +y \hat{\bm{y}}  \right) $& \\ \hline
	\multirow{3}{*}{$ B_{2u}$}		&$M_{22}^- $&$   	 y \hat{\bm{x}}+ x \hat{\bm{y}}	   	$&\multirow{3}{*}{$ k_z \left(	k_x^2-k_y^2	\right)$}\\ 
					&\multirow{1}{*}{$ M_{42}^-$}&$   x yz \hat{\bm{z}}$&	  \\ 
					&&$   z^2 \left(	y \hat{\bm{x}}+ x  \hat{\bm{y}}  \right)	 ,xy \left(	x \hat{\bm{x}} +  y\hat{\bm{y}}  \right)	$& \\\hline
	\multirow{5}{*}{$ E_{u}$}		&$M_{21}^{\pm} $&$\left[ x \hat{\bm{z}}+ z \hat{\bm{x}} , y \hat{\bm{z}} +z \hat{\bm{y}} 		\right] 	$&\multirow{5}{*}{$\left[  k_x,   k_y 		\right]  $} \\
	&$T_x,T_y$&$ [z\hat{\bm{y}}- y\hat{\bm{z}},  x\hat{\bm{z}}-z\hat{\bm{x}} ] $&\\
	&$M_{41}^{\pm} $&$\left[ z^2x \hat{\bm{z}} , z^2y \hat{\bm{z}} 		\right],   \left[ z^3 \hat{\bm{x}} , z^3 \hat{\bm{y}} 		\right], \left[ z x (x \hat{\bm{x}}+y \hat{\bm{y}} ) , y z  (x \hat{\bm{x}}+y \hat{\bm{y}} ) 		\right] 	$\\
	&\multirow{1}{*}{$M_{43}^{\pm} $}&$\left[ x \left(	x^2-y^2	\right)		 \hat{\bm{z}} , y\left(	x^2-y^2	\right)		 \hat{\bm{z}} 		\right]  , \left[ x \left(	x^2+y^2	\right)		 \hat{\bm{z}} , y\left(	x^2+y^2	\right)		 \hat{\bm{z}} 		\right]$ \\
	&&$\left[ z \left(	x^2-y^2	\right)		 \hat{\bm{x}} ,  z \left(	x^2-y^2	\right)		 \hat{\bm{y}} 	\right]  , \left[ xyz  \hat{\bm{x}} ,xyz  \hat{\bm{y}} \right]  	$& \\ \hline
		\end{tabular}
	}
		\end{table}

		\begin{table}[htbp]
		\caption{The TR even basis functions of IRs in $D_{4h}$. Basis are represented both in real space and momentum space. In the real space representation, electric multipole moments up to rank-4 are shown. The rank-5 dotriacontapole $Q_{54}^-$ is also shown for $A_{1u}$ IR.
}
		\label{eleorder}
	 {\renewcommand \arraystretch{1.35}
		\begin{tabular}{c|c|c|c}
		IR &\multicolumn{2}{c|}{Basis in real space}	& Basis in momentum space	 \\ \hline \hline
		\multirow{4}{*}{$ A_{1g}$}&$Q_{20}  $&$z^2  $&$k_z^2   	$ \\
		&$Q_{40}  $&$z^4  $&$k_z^4   	$ \\ 
		&\multirow{1}{*}{$ Q_{44}^\pm$}&$x^4-6x^2y^2+y^4  $&$k_x^4-6k_x^2k_y^2+k_y^4   	$ \\ 
		&&$x y  (x^2-y^2)  $&$k_x k_y  (k_x^2-k_y^2)   	$ \\ \hline
		$A_{2g}$	&$Q_{44}^-  $&$xy (x^2-y^2) $&$k_xk_y (k_x^2-k_y^2)  	$ \\ \hline
		\multirow{2}{*}{$ B_{1g}$}	&$Q_{22}^+  $&$ x^2-y^2	$&$k_x^2-k_y^2 	$ \\ 
			&$Q_{42}^+  $&$ (x^2-y^2)(7 z^2-r^2)	$&$( k_x^2-k_y^2 )(7k_z^2-k^2)$ \\ \hline
		\multirow{2}{*}{$ B_{2g}$} &$Q_{22}^- $&$ xy$&$k_x k_y $ \\ 
			&$Q_{42}^- $&$ xy(7 z^2-r^2)$&$k_x k_y (7 k_z^2-k^2)$ \\ \hline 
		\multirow{3}{*}{$ E_{g}$}	&$Q_{21}^{\pm}   $&$\left[ zx, yz\right]  	$&$\left[ k_zk_x, k_yk_z\right]$ \\
				&$Q_{41}^{\pm}   $&$\left[ zx(7z^2-3r^2), yz(7z^2-3r^2)\right]  	$&$\left[ k_zk_x(7k_z^2-3k^2), k_yk_z(7k_z^2-3k^2)\right]$ \\
				&$Q_{43}^{\pm}   $&$\left[ zx(x^2-3y^2), yz(x^2-3y^2)\right]  	$&$\left[ k_zk_x(k_x^2-3k_y^2), k_yk_z(k_x^2-3k_y^2)\right]$ \\ \hline
		\multirow{2}{*}{$ A_{1u}$}&\multirow{2}{*}{$(Q_{54}^- )$}&\multirow{2}{*}{$xyz \left(x^2-y^2\right)$} &$k_ x \hat{\bm{x}}+ k_y \hat{\bm{y}}+k_z \hat{\bm{z}}$	 \\ 
		&&&$ k_z \hat{\bm{z}}- k_ x \hat{\bm{x}}, k_z \hat{\bm{z}}- k_ y \hat{\bm{y}}$\\ \hline
		\multirow{2}{*}{$ A_{2u}$}&$Q_{10} $&$z$&\multirow{2}{*}{$ k_y \hat{\bm{x}}-k_x \hat{\bm{y}}  $} \\ 
		&$Q_{30} $&$z(5z^2-3r^2)$& \\ \hline 
		$ B_{1u}$	&$Q_{32}^- $&$xyz$&$   k_ x\hat{\bm{x}} -k_y \hat{\bm{y}} 	$\\ \hline
		$ B_{2u}$&$Q_{32}^+ $&$ z \left(	x^2-y^2	\right)$&$   	 k_y \hat{\bm{x}}+ k_x \hat{\bm{y}}	   	$\\ \hline
	\multirow{3}{*}{$ E_{u}$}		&$Q_{11}^{\pm} $&$\left[  x,   y  \right]  $&\multirow{3}{*}{$\left[ k_x \hat{\bm{z}} , k_y \hat{\bm{z}} 		\right]  , \left[ k_z \hat{\bm{x}} , k_z \hat{\bm{y}} 		\right]  $} \\
						&$Q_{31}^{\pm} $&$\left[  x (5z^2-r^2),   y (5z^2-r^2) \right]  $& \\
						&$Q_{33}^{\pm} $&$\left[  x (x^2-3y^2),   y (3x^2-y^2) \right]  $& \\ \hline
		\end{tabular}
	}
		\end{table}

\twocolumngrid

In the same manner, we can classify the electric multipole order. The electric multipole moment is given by
	\begin{equation}
	Q_{lm} =e \sum_{i}  r_i^l  \sqrt{\frac{4\pi}{2l+1}  } Y_{lm}^{\ast} (\theta_i ,\phi_i ). \label{elemulti}
	\end{equation}
We introduce expressions in the Cartesian coordinates as
	\begin{equation}
	\begin{split}
		Q_{lm}^+ = \frac{(-1)^m}{\sqrt{2}} \left( Q_{lm} + Q_{lm}^\ast \right), \\
		Q_{lm}^{-} = \frac{(-1)^m}{i\sqrt{2}} \left( Q_{lm} - Q_{lm}^\ast \right),
	\end{split}\label{elecubic}
	\end{equation}
for $0< l$ and $0 < m \leq l$. Table~\ref{eleorder} shows the classification of electric multipole order in tetragonal system based on the point group symmetry $D_{4h}$. For example, basis functions of $B_{1u}$ and $B_{2u}$ IRs in real space represent the electric octapole order, which has been studied in $\rm Sr_3 Ru_2 O_7$ and a bilayer Rashba system~\cite{hitomi2014,hitomi2016}. On the other hand, $A_{2u}$ and $E_{u}$ IRs correspond to the ferroelectric order, and $A_{1u}$ IR shows electric dotriacontapole order. 
Electric multipole moment is invariant under the TR operation, and therefore, the odd-parity electric multipole order parameter in $\bm{k}$-space has ``spin-dependent'' form consistent with Fermi liquid theory by Fu~\cite{Fu2015}. 
For instance, the electric octapole order is regarded as spin nematic order in $\bm{k}$-space\cite{hitomi2014}.
This is in sharp contrast to the spin-independent form of odd-parity magnetic multipole order in $\bm{k}$-space. 

\section{magnetic hexadecapole order}
Previous studies of multipole order have mainly focused on even-parity multipole formed by localized electrons~\cite{kuramoto2009}. Then, the local multipole is represented by total angular momentum multiplets, which can be systematically treated with the aid of Stevens' operator-equivalent method~\cite{kusunose2008,stevens1952,hutching1964}. With this method, even-parity multipole moment operators are recast by angular momentum operators with the use of the Wigner-Eckart theorem. On the other hand, the expectation value of odd-parity multipole moment operators vanishes when the local basis has the SI parity. Therefore, the operator-equivalent method cannot be used to evaluate odd-parity multipole moments. Hence, we should adopt local basis with mixed SI parity, which are formed by hybridization of even- and odd-parity orbitals.

\begin{figure}[htbp] 
	\centering 
	\includegraphics[width=80mm]{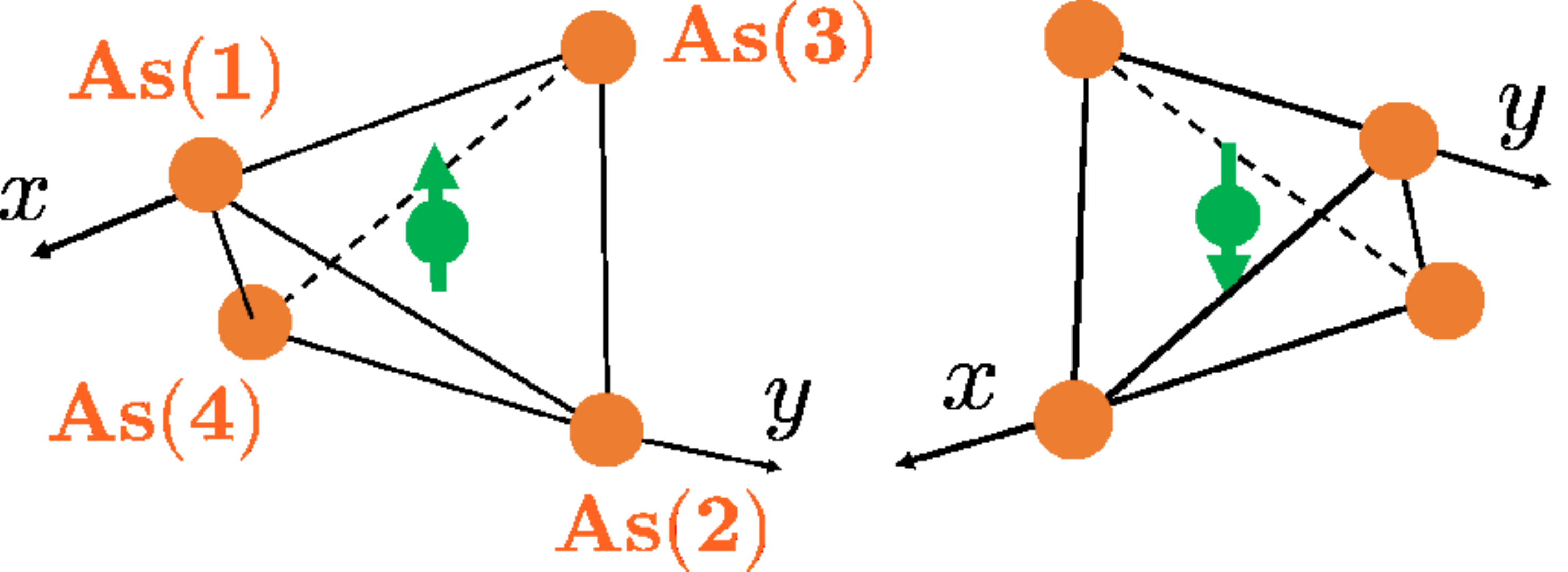} 
	\caption{(Color online)
Two nonequivalent Mn-As clusters in $\rm BaMn_2As_2$. In the AFM state, the magnetization is opposite between two clusters. Since one cluster is transformed to the other under the {\it PT} operation, the global {\it PT}  symmetry is preserved. In the left figure, As atoms surrounding Mn atoms are labeled by As(1)-As(4) for discussions. } 
	\label{cluster} 
	\end{figure}

In $\rm BaMn_2As_2$, the magnetic moment is formed mainly by Mn $d$ orbitals, and hybridization with As $p$ orbitals gives rise to the anisotropic magnetic charge distribution, namely, magnetic multipole moments. The $d$-$p$ hybridization leads to the SI parity mixing. Hence, the odd-parity magnetic multipole moments [Eq.~\eqref{magmulti}] are evaluated by calculating local magnetic multipole moments (LMMMs) of Mn-As clusters.\cite{comment1} Since the magnetic unit cell of $\rm BaMn_2As_2$ is the same as the crystal unit cell, the unit cell contains two nonequivalent Mn-As clusters shown in Figure~\ref{cluster}. Thus, we here evaluate LMMMs on the two Mn-As clusters.

In this section we consider magnetic multipole moment induced by spin angular momentum, for simplicity. With the use of the linear combination of atomic orbitals method (LCAO method), local basis is expressed by superposition of atomic orbitals on Mn and As atoms. Then, a hybridized $d$-$p$ orbital mainly consists of Mn $d$ orbital and contains As $p$ orbitals~\cite{suganobook}. With such hybrid local basis, we evaluate odd-parity magnetic multipole moments.

\subsection{Undoped $\rm BaMn_2As_2$}
Here, we calculate LMMMs of the Mn-As cluster in undoped $\rm BaMn_2As_2$. The formal valence of the $\rm Mn$ atom is $+2$ with five $3d$ electrons and the spin configuration is the completely high-spin state~\cite{johnston2011}. Thus, the orbital angular momentum quenches in the $\rm Mn$ atom. The leading odd-parity magnetic multipole moment comes from the observed $z$-component of spin magnetic moment~\cite{singh2009a,singh2009b}. Therefore, among the candidates [Eqs.~\eqref{quad} and \eqref{hexadeca}] the magnetic hexadecapole moment $M_{42,z}^+ \equiv 3 \sqrt{5}   z  \left(	x^2-y^2	\right) \hat{\bm{z}} $ is naturally the multipole order parameter of $\rm BaMn_2As_2$. 

The expectation value of $M_{42,z}^+$ is given by contributions of five electrons in the Mn-As cluster,
	\begin{align}
	\Braket{M_{42,z}^+}_{\rm L}&=\mu_{\rm B} \Braket{6 \sqrt{5} z  \left( x^2-y^2 \right) s_z}_{\rm L}  \notag \\
	&= \mu_{\rm B} \sum_{j=1}^5 \bra{\psi_{dp}^{\,j}} 6 \sqrt{5}   z  \left(x^2-y^2 \right)   \ket{\psi_{dp}^{\,j}} \Braket{s_z}_j,
	\end{align}
where the subscript $\Braket{}_{\rm L}$ indicates average on the local basis of the Mn-As cluster and we used $\hat{\bm{z}}= 2 \mu_{\rm B} s_z$. Orbital wave functions $\ket{\psi_{dp}^{\,j}}$ represent five Mn $3d$ orbitals hybridized with As $4p$ orbitals, and $\Braket{s_z}_j$ denotes spin polarization of $j$-th orbital ($j=1 - 5$). 

The two nonequivalent Mn-As clusters have the same hexadecapole moment $\Braket{M_{42,z}^+}_{\rm L}$, since both the octapole electric charge distribution $\bra{\psi_{dp}^{\,j}} 6 \sqrt{5} z  \left(x^2-y^2 \right)  \ket{\psi_{dp}^{\,j}}$ and the magnetic moment $\Braket{s_z}_j$ are opposite between the clusters. Thus, the hexadecapole moment is a ferroic order parameter as we expected. This is furthermore ensured by the symmetry; the odd-parity magnetic multipole moment operators have the even parity for the \textit{PT} symmetry and hybrid $d$-$p$ orbitals of two nonequivalent Mn-As clusters are transformed to each other under the \textit{PT} operation. Therefore, the expectation value of the hexadecapole moment $M_{42,z}^+$ is equivalent between the two clusters.

Now we evaluate the magnetic hexadecapole moment by focusing on a Mn-As cluster with $\sum_{j=1}^5 \Braket{s_z}_j >0$ without loss of generality.  
Considering quenched orbital angular momentum, we approximate the hybrid $d$-$p$ orbital by the hybrid $s$-$s$ orbital $\ket{\psi_{ss}}$ for a rough estimation. 
In the $s$-$s$ orbital, the $s$ orbitals of four $\rm As$ atoms, $\ket{s, {\rm As} (i)}$, are perturbatively hybridized with the $s$ orbital of $\rm Mn$ atom, $\ket{s,{\rm Mn}}$. The As atoms are labeled by the index $i$ (see Figure~\ref{cluster}). Denoting the hopping energy between Mn and As atoms as $-t$ and the level splitting as $\Delta_{dp}<0$, and assuming $|t/\Delta_{dp}|\ll 1$, we obtain the wave function of the hybrid orbital,   
	\begin{equation}
	\ket{\psi_{ss}} = \ket{s,{\rm Mn}} + \frac{t}{\Delta_{dp}} \sum_{i}^4 \ket{s, {\rm As}(i)}.  \label{sshybrid}
	\end{equation}
The hybridized component $\sum_i \ket{s, {\rm As} (i)}$ is not an eigenstate of the SI symmetry, indicating the local SI symmetry breaking in $\rm BaMn_2As_2$. This is an essential ingredient of the odd-parity LMMMs. We calculate the hexadecapole moment up to $O(t/\Delta_{dp})$ as 
	\begin{align}
	\Braket{M_{42,z}^+}_{\rm L}&=   \bra{\psi_{ss}} 3 \sqrt{5}   z  \left(	x^2-y^2	\right) \ket{\psi_{ss}} m_z, \notag   \\
	&=24 \sqrt{5}  \  I_{\rm EO}\frac{t}{\Delta_{dp}} m_z   + O(t^2/ \Delta_{dp}^2 ),
	\end{align}
where $I_{\rm EO} = \bra{s,{\rm As(1)}}  z (x^2-y^2) \ket{s, {\rm Mn}}$ is a matrix element of the electric octapole moment and $m_z = 2 \mu_{\rm B} \sum_{j=1}^5 \Braket{s_z}_j $ is the total spin magnetic dipole moment. 

We here adopt Slater-type orbitals~\cite{slater1930,slaterbook}, in which orbital wave functions are approximated by those of hydrogen-like atoms parametrized by effective principal quantum number $n^*$, orbital and magnetic quantum numbers $(l,m)$, and shielding factor $\alpha$:
	\begin{equation}
	\psi_{n^*, l, m, \alpha}\left( {\bm r} \right) = N r^{n^*-1} e^{-\alpha r} Y_{lm}(\hat{r}), \label{slaterorbital}
	\end{equation}
with $N$ being a normalized factor. Effective parameters $n^*$ and $\alpha$ are determined by the Slater rule~\cite{slater1930}. Real Slater-type orbitals for $l>0$ are represented by using the cubic harmonics $Z_{lm}^\pm$ instead of $Y_{lm}$. Using parameters,
\begin{equation}
\begin{split}
&(n^*,l,m,\alpha)=(3,0,0,\mr{3.52}{\ang^{-1 } }) \text{ for $\ket{s,{\rm Mn}}$},\\
&(n^*,l,m,\alpha)=(3.7,0,0,\mr{2.68}{\ang^{-1 } })	\text{ for $\ket{s, {\rm As}(1)}$},
\end{split}
\end{equation}
and position of the As(1) atom $(x, y, z)=(a/2, 0, c')$ with lattice parameters $a=\mr{4.15}{\ang}$ and $c' = \mr{1.49}{\ang}$~\cite{singh2009a,singh2009b}, 
we obtain $I_{\rm EO} = \mr{0.025}{\ang^3}$. Then, the local magnetic hexadecapole moment is evaluated as
\begin{equation}
\Braket{M_{42,z}^+}_L \simeq \mr{-0.66}{\mu_{\rm B} \ang^3} ,
\label{evaluated_hexadecapole}
\end{equation}
for $t/\Delta_{dp}=-0.1$ and $m_z = 5 \mu_{\rm B}$.

\subsection{Hole-doped $\rm BaMn_2As_2$}
Lightly hole-doped Ba$_{1-x}$K$_{x}$Mn$_2$As$_2$ shows metallic behaviors, and doping hole carriers gives the rigid band shift in the band structure~\cite{pandey2012}. Then, the magnetic structure remains to be the AFM state with a large magnetic moment $ 4.21\mu_{\rm B} $ for $x=0.05$~\cite{pandey2012,lamsal2013}. Thus, the hexadecapole moment $\Braket{M_{42,z}^+}$ is robust in the hole-doped regime. On the other hand, the hole doping changes the filling of Mn $3d$ orbitals and partially restores the orbital angular momentum, implying non-negligible effects of LS-coupling (spin-orbit coupling). This results in anisotropic distribution of magnetic charge in the $xy$-plane and induces magnetic quadrupole moment without suppressing the $\hat{z}$-collinear AFM order. 

 In a heavily hole-doped region, Ba$_{1-x}$K$_{x}$Mn$_2$As$_2$ also undergoes the ferromagnetic transition and the ferromagnetic moment is aligned in the \textit{xy}-plane~\cite{bao2012,pandey2013,pandey2015,ueland2015}. The X-ray magnetic circular dichroism experiment identified that the ferromagnetic moment arises from the As $p$ orbitals and coexists with the AFM moment of Mn atoms~\cite{ueland2015}. Although the interplay of the magnetic hexadecapole order and the ferromagnetic order would be an interesting subject, it is left for a future study. In this paper we focus on the G-type AFM state, which realizes in the lightly hole-doped region, $x < x_c \sim 0.19$~\cite{bao2012}, although we also show some numerical results beyond this doping region.

\begin{figure}[htbp] 
		\centering 
		\includegraphics[width=80mm,clip]{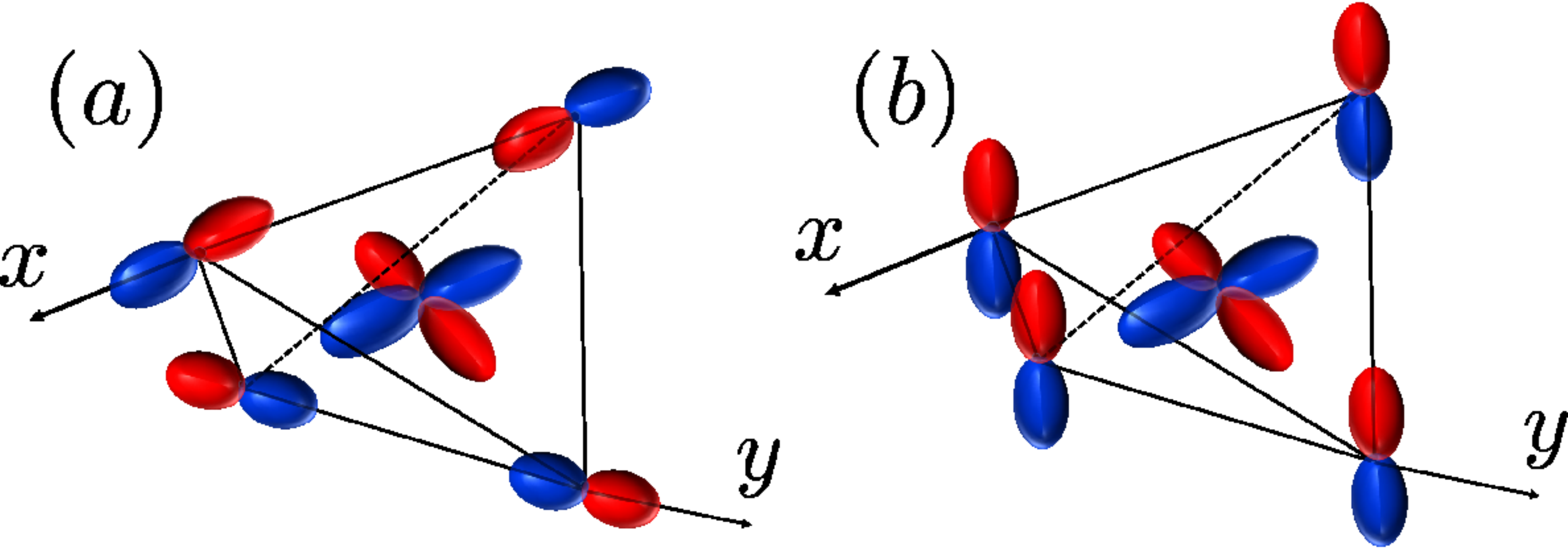} 
		\caption{(Color online)
		Sketch of $p_\alpha$ and $p_\beta$ orbitals in a Mn-As cluster. (a) The $p_\alpha$ orbital is directed to the $xy$-plane, and (b) the $p_\beta$ orbital consisting of $p_z$-orbitals extends in the $z$-direction.} 
		\label{dpcluster} 
		\end{figure}

ARPES study~\cite{zhang2016} and DFT+DMFT calculations~\cite{zingl2016} have shown that the valence band of $\rm BaMn_2As_2$ mainly consists of Mn $3d_{x^2-y^2}$ and As $4p_z$ orbitals. The doped holes occupy the hybridized $d$-$p$ orbital, whose wave function is obtained by the LCAO method, 
\begin{align}
	&\ket{\psi_{dp},\pm\frac{1}{2} } =\ket{d_{x^2-y^2},{\rm Mn}, \pm\frac{1}{2}} +	\frac{t_{\alpha}}{\Delta_{dp}}	\ket{p_\alpha, \pm\frac{1}{2}}\notag  \\
	&+ \frac{t_{\beta}}{\Delta_{dp}}	\ket{p_\beta, \pm\frac{1}{2}}  \pm \frac{i\lambda}{\Delta_1}\ket{d_{xy}, {\rm Mn}, \pm\frac{1}{2}} \notag  \\ 
	&- \frac{i\lambda}{2 \left(  \Delta_2\mp 2h		\right)	}\ket{d_{yz}, {\rm Mn}, \mp\frac{1}{2}} \pm \frac{\lambda}{2\left(	\Delta_2\mp2h	\right) }\ket{d_{zx}, {\rm Mn}, \mp \frac{1}{2}}, 
\label{dporbital}
\end{align}
with $t_{\alpha} (t_{\beta})$ being the hopping parameter between the Mn $\ket{d_{x^2-y^2}}$ and As $\ket{p_\alpha} (\ket{p_\beta})$ orbitals. Figure~\ref{dpcluster} illustrates the $p_\alpha$ and $p_\beta$ orbitals, which are given by linear combinations of $p$ orbitals of four As atoms. The orbital wave functions are explicitly written as
 	\begin{align}
	&\ket{p_\alpha} = \ket{p_x,{\rm As}(1)} - \ket{p_y,{\rm As}(2)}-\ket{p_x,{\rm As}(3)}+  \ket{p_y,{\rm As}(4)},\\
	&\ket{p_\beta} = \ket{p_z,{\rm As}(1)} + \ket{p_z,{\rm As}(2)}+\ket{p_z,{\rm As}(3)}+  \ket{p_z,{\rm As}(4)}, 
	\end{align}
which are compatible with the symmetry of the Mn $d_{x^2-y^2}$ orbital. Energy levels of As $p$ orbitals, Mn $d_{xy}$ orbital, and Mn $d_{yz}(d_{zx})$ orbital from the level of Mn $d_{x^2-y^2}$ orbital are denoted by $\Delta_{dp}$, $\Delta_1$, and $\Delta_2$, respectively. The AFM molecular field, $- h\sigma_z$ ($h>0$ for the Mn-As cluster with $m_z>0$), has been introduced for Mn $d$ orbitals, and $\lambda$ is the LS-coupling constant which is generally small in $3d$ transition metal ions. 

In the hole-doped $\rm BaMn_2As_2$, the LS-coupling induces the local magnetic quadrupole moment $M_{22}^+ = \mu_{\rm B} 2\sqrt{3} (xs_x-ys_y)$ [Eq.~\eqref{quad}] which belongs to the same IR as the magnetic hexadecapole moment $M_{42}^+$. We here calculate the expectation value of $M_{22}^+$ as follows. 
First, the contribution of one hole in the hybrid $d$-$p$ orbital $\ket{\psi_{dp},\frac{1}{2}}$ is evaluated,
	\begin{align}
	\Braket{M_{22}^+}_{\rm L}&=-\mu_{\rm B} \Braket{ 2\sqrt{3}    \left(	x s_x -y s_y	\right)  }_{\rm L}  \\
	&=  - \mu_{\rm B} 4\sqrt{3}  \frac{\lambda}{\Delta_2-2h}	\left(	I_{\alpha} \frac{t_{\alpha}}{\Delta_{dp}} + I_{\beta} \frac{t_{\beta}}{\Delta_{dp}} \right),
	\end{align}
where
	\begin{align}
	&I_{\alpha} = \bra{p_x, {\rm As}(1)}  x \ket{d_{zx},{\rm Mn}} +\bra{p_x, {\rm As}(1)} y\ket{d_{yz}, {\rm Mn}}, \\
	&I_{\beta} = \bra{p_z ,{\rm As}(1)}  x\ket{d_{zx}, {\rm Mn}} +\bra{p_z,  {\rm As}(1)}  y \ket{d_{yz}, {\rm Mn}}. 	
	\end{align}
Assuming Slater-type orbitals with effective parameters 
	\begin{equation}
	(n^*,l,m,\alpha)=(3,2,\pm1,\mr{3.52}{\ang^{-1 } }),
	\end{equation}
for the Mn $d_{yz}$ and $d_{zx}$ orbitals, and
	\begin{equation}
	(n^*,l,m,\alpha)=(3.7,1,\pm1 (0),\mr{2.68}{\ang^{-1 } }),
	\end{equation}
for the As $p$ orbitals, we obtain
	\begin{equation}
	 I_{\alpha} = \mr{-0.241}{\ang}, \,\,\, I_{\beta} = \mr{-0.0563}{\ang},
	\end{equation} 
for the lattice constant of Ba$_{1-x}$K$_{x}$Mn$_2$As$_2$ ($x=0.05$),  $a=\mr{4.16}{\ang}$ and $c' = \mr{1.49}{\ang}$~\cite{lamsal2013}. When we take $t_{\alpha}/\Delta_{dp} = t_{\beta}/\Delta_{dp} = -0.1$ and $\lambda/(\Delta_{2}-2h)=-0.01$, the magnetic quadrupole moment induced by one hole per Mn atom is estimated as,
\begin{equation}
\Braket{M_{22}^+ }_{\rm L}  =\mr{ 1.0  \times 10^{-3}}{\mu_{\rm B} \ang}. 
\end{equation}
Then, the magnetic quadrupole moment of hole-doped Ba$_{1-x}$K$_{x}$Mn$_2$As$_2$ is obtained as,
\begin{equation}
\Braket{M_{22}^+ }_{\rm L}  =\mr{x \times  5.0  \times 10^{-4}}{\mu_{\rm B} \ang}. 
\end{equation}
The magnitude of the magnetic quadrupole moment $\Braket{M_{22}^+}_{\rm L}$ is reduced by small factors $\lambda /(\Delta-2h)$ and $x$. Therefore, the magnetic hexadecapole moment remains to be the leading order parameter of hole-doped $\rm BaMn_2As_2$.

\subsection{Order parameter of odd-parity magnetic multipole order in crystals}
LMMMs specify microscopic distribution of magnetic charge around magnetic atoms or clusters, as we have studied in previous subsections. However, there are ambiguities in the definition of macroscopic odd-parity multipole moment in crystal systems. In order to avoid the ambiguity, we here introduce a unique definition by removing an irrelevant component which does not break the SI symmetry.

First, operators of multipole moment defined by Eq.~\eqref{magmulti} may depend on the origin of coordinates. Although later this ambiguity is resolved by subtracting the irrelevant component, it is convenient to choose an inversion center as the origin. Then, the magnetic unit cell is defined so that its center is the inversion center. The inversion center is no longer an inversion center in the AFM state, because the SI symmetry is spontaneously broken. However, it still remains to be an origin of the $PT$ operation preserved in the odd-parity magnetic multipole state.

\begin{figure}[htbp] 
\centering 
\includegraphics[width=80mm,clip]{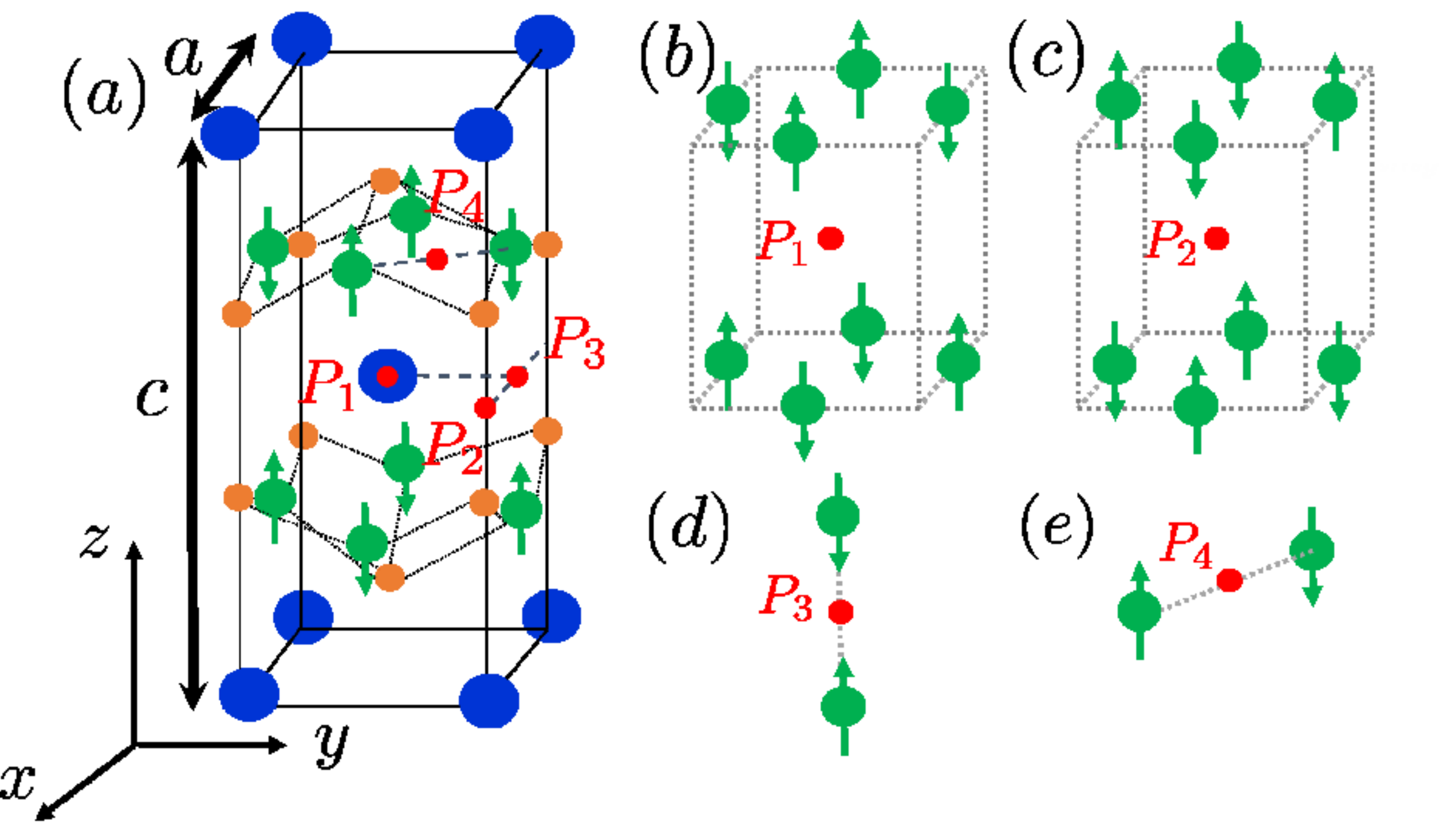} 
\caption{(Color online)
(a) A unit cell of $\rm BaMn_2As_2$. The red points show the inversion centers $P_1$, $P_2$, $P_3$, and $P_4$. (b)-(e) The magnetic unit cell corresponding to each inversion center. Configuration of neighboring Mn atoms is shown.} 
\label{maglattice} 
\end{figure}

Second, there remains an ambiguity for the choice of an inversion center and a magnetic unit cell. Actually, the crystal structure of $\rm BaMn_2As_2$ contains four nonequivalent inversion centers, namely, $P_1$, $P_2$, $P_3$, and $P_4$ in Figure~\ref{maglattice}. Coordinates of Mn atoms depend on the choice of inversion center and corresponding unit cell. The magnetic multipole moment, indeed, depends on the inversion center when it is simply defined by the expectation value of Eq.~\eqref{magmulti} in the unit cell. 
For instance, let us first choose the inversion center $P_1$. Coordinates originating from the inversion center $(X,Y,Z)$ are related to the coordinates $(x,y,z)$ used in previous subsections for Mn-As clusters; $(X,Y,Z)=(x+a/2, y, z+c/4)$ for the Mn(1) atom at $(X,Y,Z)=(a/2,0,c/4)$. Then, the expectation value of $Z(X^2-Y^2)s_z$ for electrons in the Mn-As cluster is decomposed into LMMMs and evaluated as 
	\begin{align}
	\Braket{Z (X^2-Y^2) s_z}_{\rm Mn(1)} =& 
	\Braket{z (x^2-y^2) s_z}_{\rm Mn(1)} \notag \\ 
	&+ \frac{a^2c}{16} \Braket{s_z}_{\rm Mn(1)}, \label{p1}
	\end{align} 
because the symmetry-adapted LMMM operators are only the hexadecapole moment $M_{42,z}^+$ and the dipole moment $M_{10} \propto s_z$. 
Summing up contributions from two Mn atoms in the unit cell, we obtain the multipole moment,
	\begin{equation}
	\Braket{M_{42,z}^+}_{P_1} = 2\Braket{M_{42,z}^+}_{\rm L} + \frac{3\sqrt{5} a^2 c}{4} m_z.\label{p1value}
	\end{equation}
Similarly, we obtain
	\begin{align}
	&\Braket{M_{42,z}^+}_{P_2} = 2\Braket{M_{42,z}^+}_{\rm L} - \frac{3\sqrt{5} a^2 c}{4} m_z, \label{p2value} \\
	&\Braket{M_{42,z}^+}_{P_3} = 2\Braket{M_{42,z}^+}_{\rm L},\\
	&\Braket{M_{42,z}^+}_{P_4} = 2\Braket{M_{42,z}^+}_{\rm L} ,
	\end{align}
when we choose the inversion center $P_2$, $P_3$, and $P_4$, respectively. 
We here notice that the contribution from the local magnetic dipole moment $\pm \frac{3\sqrt{5} a^2 c}{4} m_z$ causes the ambiguity. 

\begin{figure}[htbp] 
		\centering 
		\includegraphics[width=50mm]{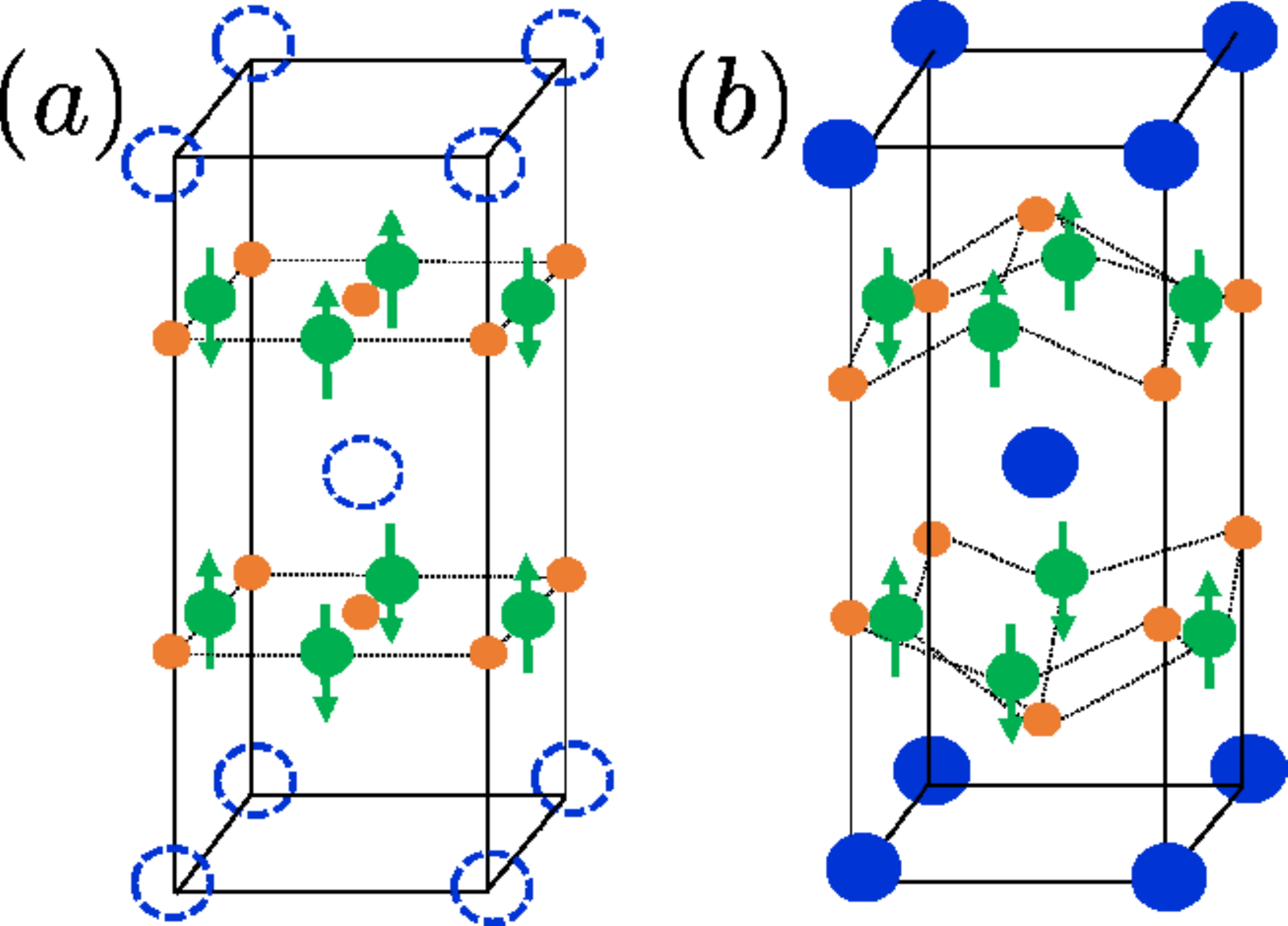} 
		\caption{(Color online)
		(a) Virtual crystal structure whose SI symmetry is recovered in the AFM state. (b) Real crystal structure of $\rm BaMn_2As_2$.} 
		\label{virtualcrystal} 
		\end{figure}

To resolve the ambiguity, we redefine the magnetic multipole moment by difference from a reference state, following procedures used for electric dipole moment~\cite{kingsmith1993,resta1994}, magnetic monopole moment~\cite{spaldin2013,thole2016}, and magnetic toroidal moment~\cite{ederer2007,spaldin2008}.
For this purpose, we consider the virtual crystal structure illustrated in Figure~\ref{virtualcrystal}. In the virtual crystal structure [Figure~\ref{virtualcrystal}(a)], the As atoms lie in the same plane as Mn atoms and Ba atoms have been removed. Then, the $D_{4h}$ symmetry is preserved even in the AFM state, since the Mn atoms are inversion centers. 
However, the magnetic hexadecapole moment defined by Eq.~\eqref{magmulti} remains finite for the inversion center $P_1$ and $P_2$ due to the irrelevant terms, $\pm \frac{3\sqrt{5} a^2 c}{4} m_z$.  
Thus, we define the order parameter of odd-parity magnetic multipole order $\Braket{M_{lm}}$ by subtracting the irrelevant component,
	\begin{equation}
	\Braket{M_{lm}} =\Braket{M_{lm}}_\Gamma - \Braket{M_{lm}}_{\Gamma}^{0}, \label{magop}
	\end{equation}
where $\Gamma$ denotes an inversion center and $\Braket{}^0_\Gamma$ indicates the expectation value in the virtual crystal structure, namely, the reference state. 
Although the multipole moment $\Braket{M_{lm}}_{\Gamma}^0$ in the reference state depends on an inversion center, the odd-parity magnetic multipole moment defined by difference from the reference state is unique in the sense that it is independent of the choice of inversion center and unit cell. 

The local magnetic hexadecapole moment vanishes in the virtual crystal, namely, $\Braket{M_{42,z}^+}_{\rm L}^0 =0$, since the Mn-As clusters preserve the local SI symmetry. In other words, the macroscopic magnetic hexadecapole moment defined above is given by the LMMM,
\begin{equation}
	\Braket{M_{42,z}^+} =2 \Braket{M_{42,z}^+}_{\rm L}. \label{hexadecaorder}
\end{equation}
Similarly, we obtain
\begin{equation}
	\Braket{M_{22}^+} =2 \Braket{M_{22}^+}_{\rm L}, \label{quadorder}
\end{equation}
for the magnetic quadrupole moment. Thus, the macroscopic magnetic multipole moment in $\rm BaMn_2 As_2$ is given by the LMMMs investigated in Secs.~IIIA and IIIB. It is again stressed that the local SI symmetry breaking in the crystal structure plays an essential role for the odd-parity magnetic multipole order.

The procedure used in this subsection can be applied to not only $\rm BaMn_2As_2$ but also various odd-parity magnetic multipole states. First, a magnetic and centrosymmetric crystal structure is considered as a reference state. Second, an irrelevant component which is finite in the centrosymmetric state is evaluated. Then, the odd-parity magnetic multipole moment in real crystals is uniquely defined by difference from the reference state. 
The reference state is not uniquely determined in general. However, it is reasonable to consider the virtual structure in Figure~\ref{virtualcrystal} for $\rm BaMn_2As_2$ as a reference state which restores the local SI symmetry of magnetic sites. 
Using this framework, we are able to estimate odd-parity magnetic multipole moment more precisely by first-principles calculations~\cite{spaldin2013,thole2016,bultmark2009,cricchio2010,suzuki2017}. The first principles study of Ba$_{1-x}$K$_x$Mn$_2$As$_2$ is an important future work. 

 For calculations of the multipole moment, additional care is needed for the multivalued problem~\cite{kingsmith1993,ederer2007,thole2016}. When evaluating electric dipole moment by using the Berry phase formulation~\cite{kingsmith1993}, we may obtain the electric dipole moment with the arbitrariness of $ne\bm{R}$, where $n$ is an integer and $\bm{R}$ is the minimal lattice vector along the polarization axis. The physically meaningful dipole moment should be smaller than the arbitrary term. 
Similar multivalued problem may also occur in calculations of higher-order multipole moment. The arbitrary term of magnetic hexadecapole moment, namely, the quantum unit of magnetic hexadecapole moment $\Delta M_{42,z}^+$ is roughly evaluated as
	\begin{equation}
	\Delta M_{42,z}^+ \sim a^2 c \, m_z,
	\end{equation}
which is in the same order as the irrelevant terms $\pm \frac{3\sqrt{5} a^2 c}{4} m_z$ in Eqs.~\eqref{p1value} and \eqref{p2value}. Our evaluation of the magnetic hexadecapole moment $\Braket{M_{42,z}^+} \sim \mr{1}{\mu_{\rm B} \ang^3}$ [Eqs.~\eqref{evaluated_hexadecapole} and \eqref{hexadecaorder}] is much smaller than the quantum unit $\Delta M_{42,z}^+ \sim \mr{10^2}{\mu_{\rm B} \ang^3}$, and therefore our calculation does not suffer the multivalued problem.

\section{Effective model}
In the following part of this paper, we show characteristic properties induced by odd-parity magnetic multipole order. For this purpose, we introduce a tight-binding Hamiltonian for the valence band of $\rm BaMn_2 As_2$ mainly consisting of Mn $d_{x^2-y^2}$ orbital~\cite{zhang2016,zingl2016}.

By projecting the five-orbital model to the valence band (Appendix A), the effective Hamiltonian is obtained as 
\begin{align}
	\mathcal{H} &= \mathcal{H}_{\rm hop} + \mathcal{H}_{\rm ASOC}+ \mathcal{H}_{\rm AFM}  \label{efhamiltionian0} 
	=\sum_{\bm{k}} \bm{c}_{\bm{k}}^\dagger H(\bm{k})  \bm{c}_{\bm{k}}, \\
	H(\bm{k}) &\scalebox{0.9}{$\displaystyle =\begin{pmatrix}
			\epsilon(\bm{k}) + \left[ \bm{g}_{\rm A}(\bm{k}) -\bm{h}_{\rm A}  \right] \cdot \bm{\sigma}& V_{\rm AB} (\bm{k})\\
			 V_{\rm AB} (\bm{k})&\epsilon(\bm{k})   +\left[ \bm{g}_{\rm B}(\bm{k}) -\bm{h}_{\rm B}  \right] \cdot \bm{\sigma}
			 \end{pmatrix} $},
\end{align}
where $\bm{\sigma}= (\sigma_x, \sigma_y, \sigma_z)$ is the Pauli matrix and $\bm{c}_{\bm{k}} = \left(c_{\bm{k}, \rm A, +}, c_{\bm{k}, \rm A, -}, c_{\bm{k}, \rm B, +}, c_{\bm{k}, \rm B, -}	\right)^{ T} $ is a vector representation of annihilation operators labeled by momentum $\bm{k}$, sublattice index $\tau=A, B$, and spin $\sigma=\pm $. 
The kinetic energy term is given by
\begin{align}
\mathcal{H}_{\rm hop} &= \sum_{\bm{k}, \tau, \sigma} \epsilon (\bm{k}) c_{\bm{k}, \tau, \sigma}^\dagger c_{\bm{k}, \tau, \sigma} \notag \\
&+\sum_{\bm{k}, \sigma}\left(  V_{\rm AB} (\bm{k}) c_{\bm{k}, \rm A, \sigma}^\dagger c_{\bm{k}, \rm B, \sigma} + h.c.		\right),
\end{align}
where 
\begin{align}
&\epsilon (\bm{k})= -2t_1 \left(	\cos{ k_x  } +\cos{ k_y }	\right)-8 t_2 \cos{\frac{k_x}{2} } \cos{\frac{k_y}{2} }\cos{\frac{k_z}{2} },\\
&V_{\rm AB }(\bm{k})= -4\tilde{t}_1  	\cos{ \frac{k_x}{2}  } \cos{ \frac{k_y}{2} } -2 \tilde{t}_2 \cos{\frac{k_z}{2} },
\end{align}
are intra-sublattice and inter-sublattice hopping energy, respectively. The G-type AFM structure of $\rm BaMn_2As_2$ is taken into account by the molecular field term $\mathcal{H}_{\rm AFM}$. Since the magnetic moment is parallel to the $z$-axis and changes its sign between the A and B sublattices, the AFM molecular field is given by  $\bm{h}_{\rm A} = h\hat{z}$ and $\bm{h}_{\rm B} = -h\hat{z}$.

The $\bm{k}$ dependent Zeeman terms $\bm{g}_{\rm A(B)} (\bm{k} ) \cdot \bm{\sigma}$ originate from the LS-coupling and the inter-orbital hybridization between the Mn $d_{x^2-y^2}$ orbital and other Mn $d$ orbitals, as we show the derivation from the five-orbital model in Appendix A. The $PT$ symmetry preserved in the AFM state ensures the staggered structure $\bm{g}_{\rm A} (\bm{k})= -\bm{g}_{\rm B} (\bm{k}) \equiv \bm{g} (\bm{k})$. The $g$-vector $\bm{g}(\bm{k})$ is decomposed into the odd-parity and even-parity parts, $\bm{g}(\bm{k}) = \bm{g}'(\bm{k}) +\bm{g}''(\bm{k})$. The odd-parity component represents the ASOC term by 
\begin{equation}
\bm{g}'(\bm{k}) = \begin{pmatrix}
			\alpha_1  \sin{k_y} + \alpha_2 \cos{\frac{k_x}{2} } \sin{\frac{k_y}{2} }\cos{\frac{k_z}{2} }\\
			\alpha_1  \sin{k_x} + \alpha_2 \sin{\frac{k_x}{2} } \cos{\frac{k_y}{2} }\cos{\frac{k_z}{2} }\\
			\alpha_3 \sin{\frac{k_x}{2} } \sin{\frac{k_y}{2} }\sin{\frac{k_z}{2} }
			\end{pmatrix}. 
\end{equation}
This term arises from the local SI symmetry breaking of Mn atoms, and therefore, all the coefficients $\alpha_1$, $\alpha_2$, and $\alpha_3$ are finite in both paramagnetic and magnetic hexadecapole states. In contrast, the additional component $\bm{g}''(\bm{k})$ denotes an even-parity spin-orbit coupling, called as symmetric spin-orbit coupling (SSOC). The derivation from the five-orbital model gives the expression (see Appendix A),
\begin{equation}
\bm{g}''(\bm{k}) = \begin{pmatrix}
			\beta  \sin{\frac{k_x}{2} } \cos{\frac{k_y}{2} }\sin{\frac{k_z}{2} }\\
			-\beta   \cos{\frac{k_x}{2} } \sin{\frac{k_y}{2} }\sin{\frac{k_z}{2} }\\
			0
			\end{pmatrix}.
\end{equation}
The SSOC term breaks the TR symmetry, although it breaks neither the local nor global SI symmetry. Therefore, the SSOC term disappears in the paramagnetic state. The broken TR symmetry by the AFM order gives rise to the SSOC term. 

Diagonalizing the Bloch Hamiltonian $H(\bm{k})$, we obtain the energy spectrum
\begin{equation}
E_{\bm{k}} = \epsilon (\bm{k}) \pm \sqrt{V_{\rm AB}(\bm{k})^2 + \left|\bm{g} (\bm{k})-h\hat{z} \right|^2 }, \label{energyspectrum}
\end{equation}
with double degeneracy protected by the $PT$ symmetry. In the undoped system, the Fermi level lies in the gap of the two bands. Then, the system shows insulating behaviors. Doping hole carriers lowers the Fermi level without reconstruction of the band structure~\cite{pandey2012}. Then, the partially filled valence band leads to metallic behaviors.   

In the following sections, we investigate electromagnetic responses resulting from the SI symmetry breaking. Then, the SSOC term does not play an important role since it does not break local or global SI symmetry, as discussed in Appendix A. Thus, we set $\beta=0$ for simplicity, and assume the parameters, $t_1 = -0.1$, $t_2=-0.05$, $\tilde{t}_1=0.05$, $\tilde{t}_2=0.01$ for the kinetic energy term, $h=1$ for the AFM molecular field, and $\alpha_1 = -0.005$, $\alpha_2 =0.001$, $\alpha_3 = 0.01$ for the ASOC term, unless mentioned otherwise. The interlayer coupling is moderate in BaMn$_2$As$_2$ compared with a related quasi-two-dimensional compound LaMnAsO~\cite{zingl2016}. Thus, moderate interlayer hopping integrals are assumed. 
We adopt the unit for the lattice parameter, $a=c=1$. 

\section{Magnetoelectric effect}
\subsection{Uniform magnetoelectric effect}
In the previous sections, the AFM state of $\rm BaMn_2 As_2$ has been identified as an odd-parity magnetic multipole state, where the SI and TR symmetry are broken whereas the combined $PT$ symmetry is preserved. Then, the ME coupling is allowed in a free energy expansion in accordance with group-theoretical discussions~\cite{dzyaloshinskii1959}. The resulting ME effect, $\bm{M}=  \hat{\alpha} \bm{E}$, that is, the electric field-induced magnetization has been observed in experiments~\cite{astrov1960}. The symmetry argument tells us that the ME response is attributed to rank-2 magnetic multipole orders listed in Table \ref{d4h_magmulti}. Decomposing the ME tensor $\hat{\alpha} = \left(\alpha_{\mu \nu}\right)$ into isotropic, antisymmetric, and traceless symmetric terms, we have 
	\begin{align}
	\hat{\alpha} &=  \frac{1}{3} \left( \rm{Tr}  \hat{\alpha} \right)  \hat{1} + \frac{1}{2}\left( \hat{\alpha} -\hat{\alpha}^T \right) 
	+ \left[ \frac{1}{2}\left( \hat{\alpha} + \hat{\alpha}^T \right) -  \frac{1}{3} \left( \rm{Tr}  \hat{\alpha} \right) \hat{1} \right],
	\end{align}
corresponding to magnetic monopole moment, $\sum_i x_i \hat{\bm{x}}_i$, magnetic toroidal dipole moment, $\sum_{j,k} \epsilon_{ijk}x_j \hat{\bm{x}}_k$, and magnetic quadrupole moment, $x_i \hat{\bm{x}}_j +x_j \hat{\bm{x}}_i$ and $x_i \hat{\bm{x}}_i-x_j \hat{\bm{x}}_j$ for $i\neq j$. In accordance with the symmetry of magnetic quadrupole moment, $M_{22}^+ \propto  x\hat{\bm{x}}-  y\hat{\bm{y}}$, the ME effect characterized by the ME tensor 
\begin{align}
\hat{\alpha}=
\begin{pmatrix}
\alpha & 0 & 0 \\
0 & -\alpha & 0 \\
0 & 0 & 0 \\
\end{pmatrix},
\end{align}
is allowed in BaMn$_2$As$_2$. 

To demonstrate the ME effect, we calculate the ME coefficient by Kubo formula, 
\begin{equation}
\alpha_{\mu \nu } = \frac{e g\mu_{\rm B} \hbar}{2i N} \sum_{\bm{k}, p,q} \frac{ \left[ \sigma_\mu (\bm{k}) \right]_{pq} \left[ v_\nu (\bm{k})\right]_{qp} }{E_{p} (\bm{k})-E_{q} (\bm{k}) +i\delta} \frac{f( E_{p})  -f( E_{q} )  }{ E_{p} (\bm{k})-E_{q} (\bm{k}) }, \label{magneto}
\end{equation}
where $p$ and $q$ label the band indices, 
$N$ is the number of unit cell, $\delta$ is a scattering rate, and $f(E)$ is the Fermi distribution function. $\left[ \sigma_\mu (\bm{k}) \right]_{pq}$ and $\left[ v_\nu (\bm{k}) \right]_{qp}$ are respectively the band representation of spin operator $\sigma_\mu$ and velocity operator $v_\nu(\bm{k}) = \partial H(\bm{k}) / \partial k_\nu$.

\begin{figure}[htbp]
\centering
\includegraphics[width=80mm,clip]{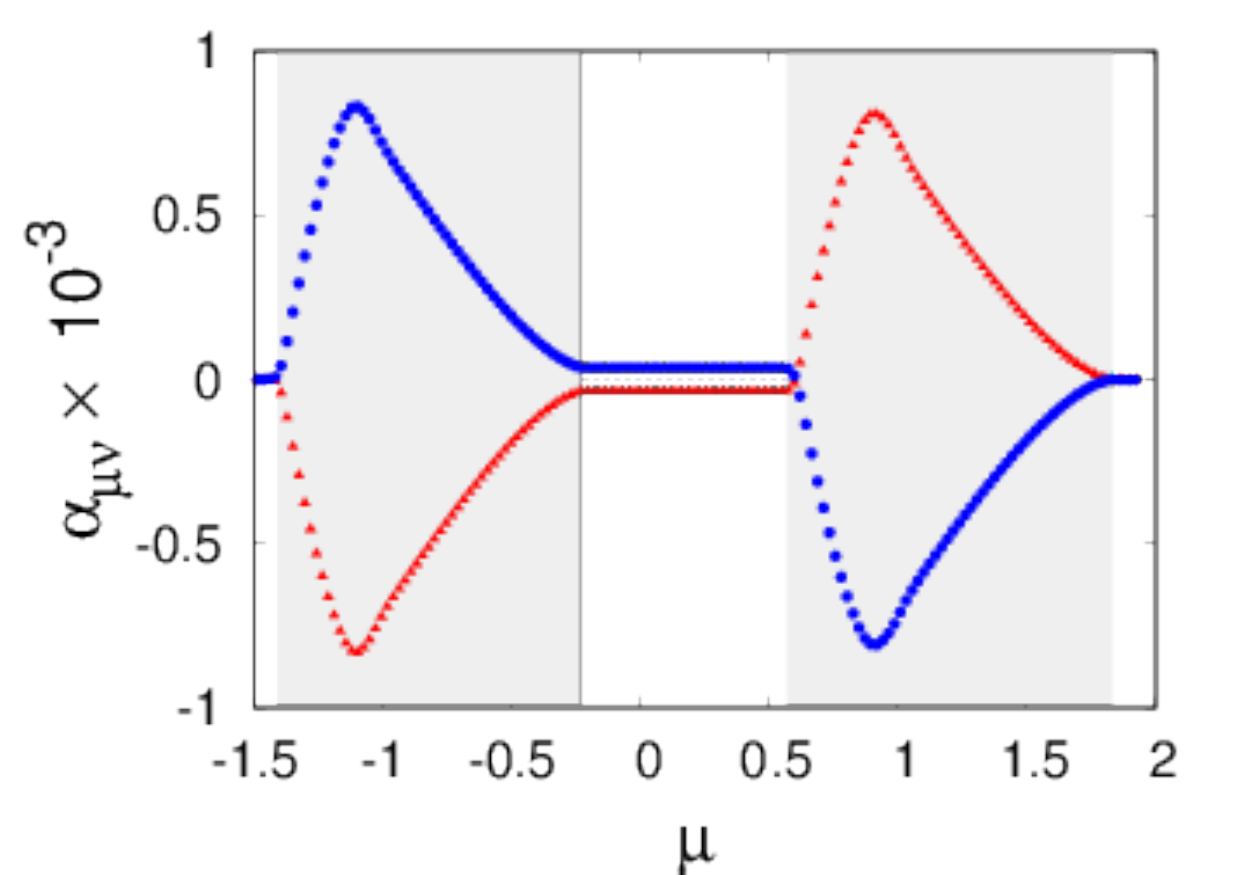}
\caption{(Color online)
ME coefficients $\alpha_{xx}$ (red triangles) and $\alpha_{yy}$ (blue circles) as a function of the chemical potential $\mu$. We assume the temperature $T=0.01$ and the scattering rate $\delta=0.01$, and choose the unit $eg\mu_{\rm B}\hbar/2 =1$. }
\label{magnetoplot}
\end{figure}

We plot the ME coefficient as a function of the chemical potential $\mu$ in Figure~\ref{magnetoplot}. 
Our numerical result is consistent with the symmetry argument. Only the ME coefficients $\alpha_{xx} = - \alpha_{yy}$ are finite, corresponding to the magnetic quadrupole moment $M_{22}^+$. 
Dark background in the figure represents the metallic region where the chemical potential lies in the valence band or conduction band. Otherwise, the chemical potential lies in the gap, and the system is insulating. Interestingly, the ME effect is significantly enhanced in the metallic region. 
The magnitude of the magnetoelectric coupling in the insulating phase ($| \alpha_{xx} |= | \alpha_{yy} | \sim 10^{-4}$) corresponds to $\sim \mr{10^{-3}}{ps \cdot m^{-1}}$ when we take $|t_1|=\mr{100}{meV}$. This magnetoelectricity is much smaller than that of the prototypical magnetoelectric material $\rm Cr_2 O_3$ ($|\alpha_{\perp}|=|\alpha_{xx}|=|\alpha_{yy}| \sim \mr{10^{-1}}{ps \cdot m^{-1}}$, $|\alpha_{\parallel}|=|\alpha_{zz}|\sim \mr{1}{ps \cdot m^{-1}}$)~\cite{folen1961}, because only a small magnetic quadrupole moment is induced by the LS-coupling term. However, precise estimation of the magnetoelectric coupling requires more elaborate works. For instance, calculations based on the multiorbital model, estimation of orbital magnetoelectricity~\cite{malashevich2010,scaramucci2012}, and DFT calculations are desired. 

\subsection{Antiferromagnetic Edelstein effect}
Next, we show the AFM Edelstein effect, namely, the AFM spin polarization induced by the electric current. 
This characteristic response of locally noncentrosymmetric systems~\cite{yanase2014} is attracting 
recent interest for application to antiferromagnetic spintronics~\cite{zelezny2014,wadley2016,gomonay2016}.

The operator of AFM spin moment is defined as $\sigma_\mu^{\rm AF} = \sigma_\mu \tau_z$ with the Pauli matrix $\bm{\tau}$ acting on the sublattice space.
The ASOC term is uniform between sublattices when it is represented by the AFM spin operator, 
\begin{align}
\mathcal{H}_{\rm ASOC} &= \sum_{\bm{k}, \tau, \sigma, \sigma'} {\bm g}(\bm{k}) \cdot {\bm \sigma}^{\rm AF} c_{\bm{k}, \tau, \sigma}^\dagger c_{\bm{k}, \tau, \sigma'}.  
\end{align}
Hence, the AFM spin-momentum locking occurs in locally noncentrosymmetric systems, that is analogous to the spin-momentum locking in globally noncentrosymmetric systems. 
The above representation of the ASOC term indicates the staggered ME effect represented by 
\begin{align}
\bm{M}^{\rm AF}=  \hat{\alpha}^{\rm AF} \bm{E}, 
\label{AFM_Edelstein}
\end{align} 
in analogy to the Edelstein effect~\cite{Edelstein}, that is, the spin polarization due to the current-induced shift of Fermi surface. Since the ASOC term contains an in-plane component, $\propto k_x \sigma_y^{\rm AF} +k_y \sigma_x^{\rm AF} $, which is a basis function of the totally-symmetric $A_{1g}$ IR of the $D_{4h}$ point group, we have finite staggered ME coefficients, $\alpha_{yx}^{\rm AF} = \alpha_{xy}^{\rm AF}$. The ASOC term is derived from only the local SI symmetry breaking and does not require the TR symmetry breaking. Therefore, the staggered ME effect occurs in both paramagnetic state and AFM state. 
 
\begin{figure}[htbp]
\centering
\includegraphics[width=90mm,clip]{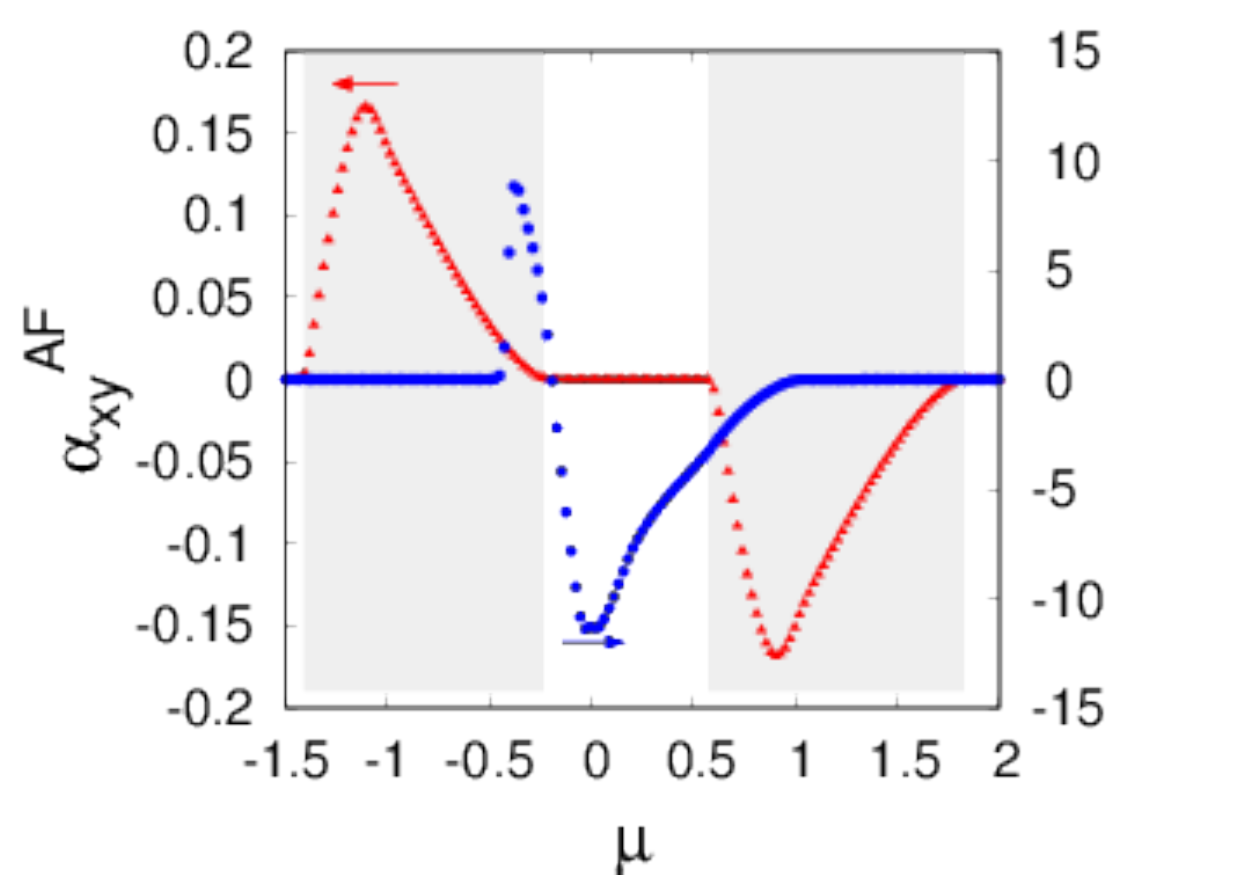}
\caption{(Color online)
Staggered ME coefficient $\alpha_{xy}^{\rm AF}$ ($=\alpha_{yx}^{\rm AF}$) as a function of the chemical potential $\mu$ in the AFM state ($h=1$, red triangles) and in the paramagnetic state ($h=0$, blue circles). The shaded area indicates the metallic region for $h=1$. 
We assume $T=0.01$ and $\delta=0.01$ and adopt the unit $eg\mu_{\rm B}\hbar/2 =1$.}
\label{antimagnetoplot}
\end{figure}

Replacing $\sigma_\mu $ with $\sigma_\mu^{\rm AF}$ in Eq.~\eqref{magneto}, we calculate the staggered ME coefficient $\alpha_{\mu\nu}^{\rm AF}$. Figure~\ref{antimagnetoplot} shows numerical results of $\alpha_{yx}^{\rm AF} = \alpha_{xy}^{\rm AF}$ in the AFM state ($h=1$) and the paramagnetic state ($h=0$). 
Although the staggered ME effect is caused by the in-plane component of the ASOC term, $g'_x({\bm k})$ and $g'_y({\bm k})$, this component is suppressed in the AFM state (see Appendix A) since the spin polarization along the $z$-axis suppresses the spin-flipping process. Thus, we assume the in-plane components in the paramagnetic state, $\alpha_1$(para)$=-0.05$ and $\alpha_2$(para)$=0.01$, which are larger than those in the AFM state. 
Indeed, the staggered ME coefficient is smaller in the AFM state than in the paramagnetic state. 

In contrast to the uniform ME effect, the staggered ME effect is essentially induced by the electric current. The shift of Fermi surface under the current results in the AFM spin polarization, like in the Edelstein effect~\cite{Edelstein}. Thus, we call Eq.~\eqref{AFM_Edelstein} the AFM Edelstein effect. 

The difference between the uniform ME response and staggered ME response comes from the $PT$ parity of spin operators. The $PT$ parity is even for the AFM spin moment $\bm{\sigma}^{\rm AF}$, while the uniform spin operator $\bm{\sigma}$ is $PT$ odd. Since the velocity operator $\bm{v}$ has even $PT$ parity, the ME coefficient $\alpha_{\mu\nu} $ is purely determined by interband effects, whereas the staggered ME coefficient $\alpha^{\rm AF}_{\mu\nu}$ by intraband effects. Thus,  $\alpha_{\mu\nu} \propto \tau^0$ while $\alpha^{\rm AF}_{\mu\nu} \propto \tau^1$ with respect to the lifetime of quasiparticles, indicating the {\it electric field-induced} uniform ME effect and the {\it electric current-induced} AFM Edelstein effect. Indeed, the latter does not occur in the insulating state. In Appendix B, we prove the lemma for Kubo formula supporting these discussions. 

\begin{figure}[htbp]
\centering
\includegraphics[width=80mm,clip]{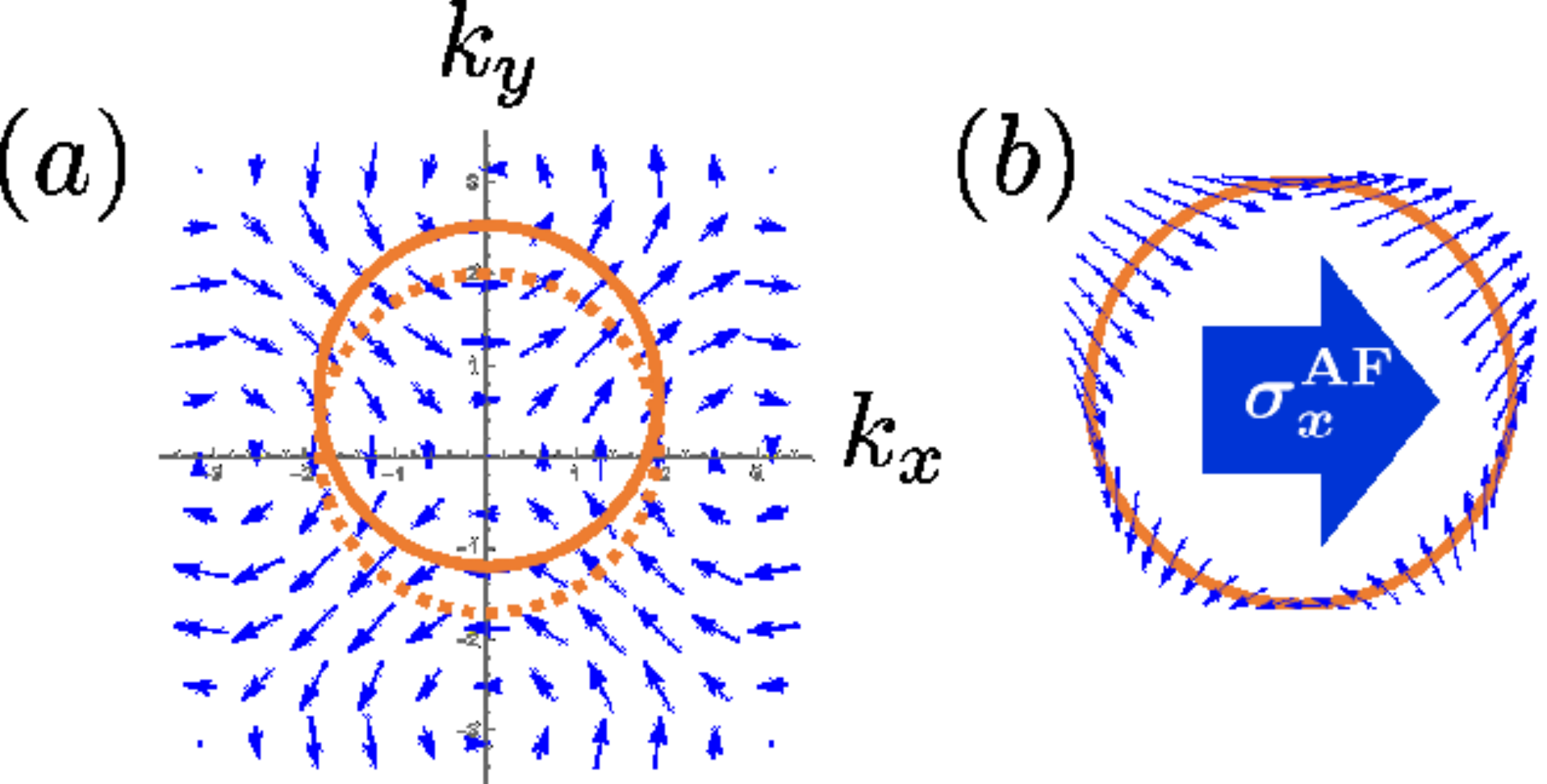}
\caption{(Color online)
(a) The in-plane component of $g$-vector [$\bm{g}'_x({\bm k})$, $\bm{g}'_y({\bm k})$] in the ASOC term. The momentum dependence on the $k_z =0$ plane is shown by arrows.  The electric field along the $y$-direction shifts the Fermi surface (circle with a solid line). (b) The ${\bm k}$-dependent AFM spin polarization on the Fermi surface. Summation for the momentum leads to a macroscopic AFM spin polarization in the $x$-direction.
}
\label{basalanti}
\end{figure}

Here, we show a simplified expression of $\alpha_{xy}^{\rm AF}$. The matrix element of the velocity operator is obtained as
\begin{align}
\left[	v_\mu (\bm{k})\right]_{pq} &= \frac{\partial E_p(\bm{k})}{\partial k_\mu} \delta_{pq} \notag \\
&+ \left( E_p(\bm{k})-E_q(\bm{k}) \right) \Braket{ u_p (\bm{k}) | \frac{\partial  u_q (\bm{k}) }{\partial k_\mu}	},
\end{align}
where $\ket{u_p (\bm{k})}$ denotes Bloch states satisfying $H(\bm{k}) \ket{u_p(\bm{k})} =E_{p}(\bm{k}) \ket{u_p(\bm{k})}$. Summing up intraband contributions, we obtain 
\begin{align}
\alpha_{xy}^{\rm AF} &\simeq \frac{e g\mu_{\rm B} \hbar}{i N} \sum_{\bm{k},p} \frac{ \left[ \sigma_x^{\rm AF} (\bm{k})\right]_{pp}}{i\delta} \frac{\partial E_p(\bm{k})}{\partial k_y}\frac{ \partial f( E) }{\partial  E }\Big|_{E_p }  \notag \\
&= \frac{-e g\mu_{\rm B} \hbar}{\delta N} \sum_{\bm{k},p} 	 \left[ \sigma_x^{\rm AF} (\bm{k})\right]_{pp}   \frac{ \partial f( E_p ) }{\partial k_y}.
\end{align}
Because $\tau = 1/\delta$, we confirm $\alpha^{\rm AF}_{xy} \propto \tau^1$. 
At low temperatures, the staggered ME coefficient $\alpha_{xy}^{\rm AF}$ is determined by quasiparticles near the Fermi surface. Since the AFM spin moment is locked to momentum due to the ASOC term, the deformation of Fermi surface represented by $\partial f (E) /\partial k_y $ gives rise to the finite AFM moment $M_x^{\rm AF} = \Braket{{\rm Tr} \sum_{\bm{k}} \sigma_{x}^{\rm AF} (\bm{k})}$, 
as schematically shown in Figure~\ref{basalanti}.

The AFM Edelstein effect enables electrical switching of AFM domain~\cite{zelezny2014,wadley2016}, 
pointing to the AFM spintronics~\cite{gomonay2016}. However, the seemingly AFM structure is classified into 
the odd-parity magnetic multipole. In other words, the ``AFM domain'' switched by the electric current is, indeed, the domain of {\it ferroic} odd-parity magnetic multipole from the viewpoint of multipole physics. 
Although it may be expected that the magnetic hexadecapole moment of $\rm Ba Mn_2 As_2$ can be switched 
by injecting an electric current, it is unlikely at least in the linear response region. 
The effective AFM Zeeman field driven by electric field is confined to the $xy$-plane and the AFM Edelstein effect cannot switch the $z$-collinear AFM domains of $\rm BaMn_2As_2$. However, an uniaxial strain along the in-plane direction reduces the site symmetry of Mn atoms and accordingly induces another AFM Edelstein effect characterized by a finite coefficient $\alpha_{z z}^{\rm AF}$. Then, the magnetic hexadecapole moment may be switched by the electric current along the $z$-direction: $M_{42}^+ >0 \leftrightarrow M_{42}^+<0$. Furthermore, the electric current along the $z$-direction induces the strain field in the $xy$-plane, as we show in Sec.~VI. Therefore, the nonlinear effect of the electric current gives the effective AFM Zeeman field and may switch the magnetic hexadecapole domain. 

\section{Current-induced nematicity}

We here show a counter-intuitive response in the metallic magnetic multipole state. 
The electric current along the {\it z}-axis induces the nematicity in the {\it xy}-plane. 

\begin{figure}[htbp]
\centering
\includegraphics[width=80mm,clip]{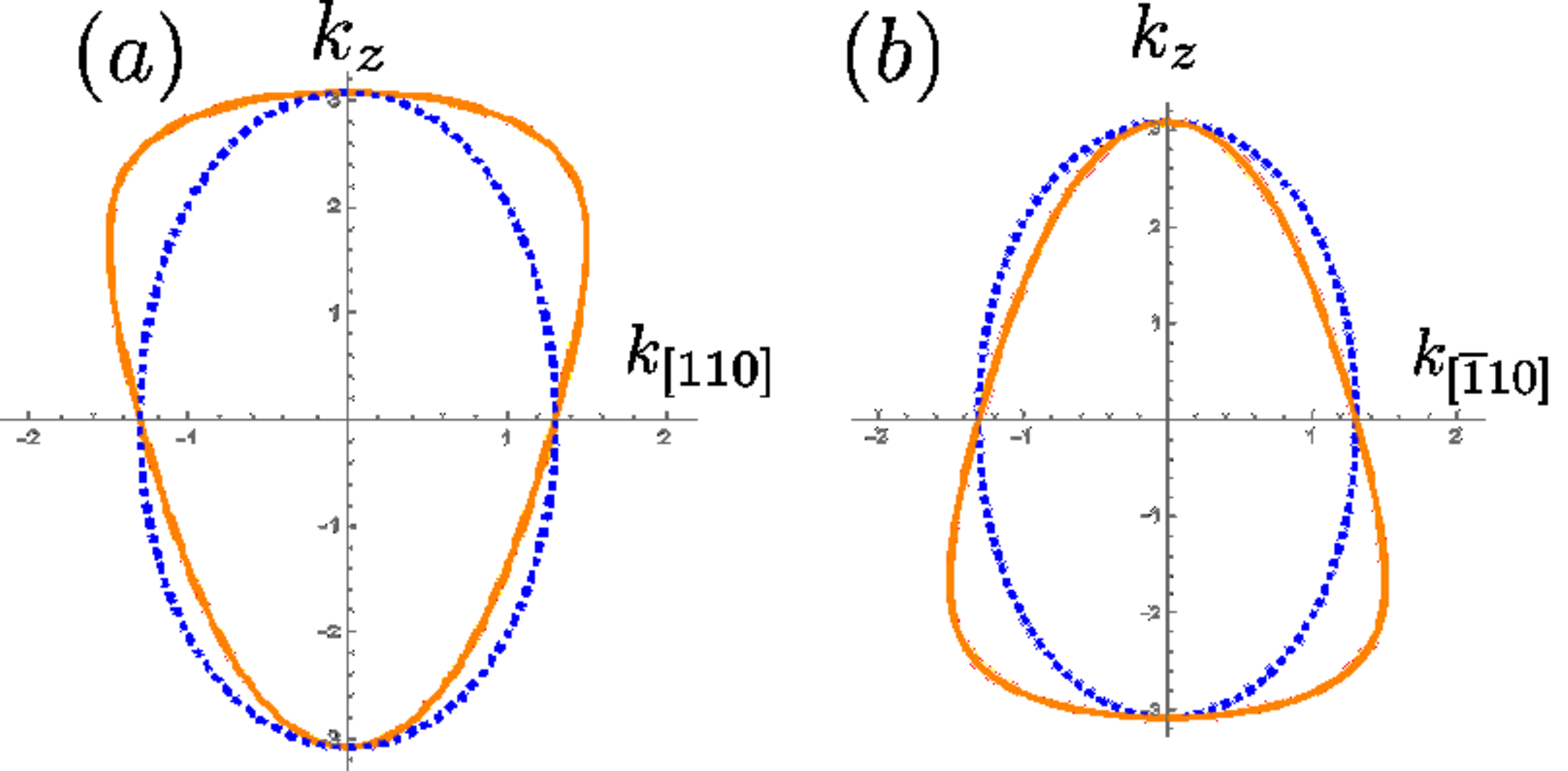}
\caption{(Color online)
The tetrahedral modulation of Fermi surface is shown with the solid lines. Fermi surfaces in the (a) $k_{[110]}-k_z$ plane and (b) $k_{[\overline{1}10]}-k_z$ plane are shown. $k_{[110]}$ and $k_{[\overline{1}10]}$ are momentum along the $[110]$ and $[\overline{1}10]$ direction, respectively. The chemical potential is set to $\mu=-0.8$ and a large ASOC $\alpha_3=0.3$ is assumed for emphasizing the tetrahedral modulation. Fermi surfaces for $\alpha_3=0$ are plotted with the dashed lines for a comparison.}
\label{surfaceplot}
\end{figure}

As we show in Table \ref{magorder}, the order parameter of odd-parity magnetic multipole order represented in $\bm{k}$-space indicates spin-independent asymmetric modulation of the band structure, which has been demonstrated in several models~\cite{hayami2014,sumita2016,yanase2014,sumita2017}. In $\rm BaMn_2 As_2$, the order parameter of the $B_{1u}$ IR is $k_x k_y k_z $ in $\bm{k}$-space. The corresponding cubic asymmetry in the energy spectrum results from the coupling of the AFM molecular field and the $z$-axis component of the ASOC term, which is indeed, $ -h g'_3 (\bm{k}) \propto k_x k_y k_z$, in the long-wavelength limit. This term induces a tetrahedral modulation of Fermi surfaces as shown in Figure~\ref{surfaceplot}. The same modulation also arises from the coupling between the ASOC and SSOC terms, $\bm{g}'(\bm{k}) \cdot \bm{g}''(\bm{k})$, although this term is negligible as we discussed in Appendix A. 

The electric current along the {\it z}-axis induces finite expectation value of $k_z$. Then, the tetrahedral modulation leads to $k_x k_y k_z \rightarrow k_x k_y \langle k_z \rangle$, indicating the nematicity in the {\it xy}-plane resulting from the nematic modulation of Fermi surface. This is an intuitive explanation of the current-induced nematicity shown below.

\begin{figure}[htbp]
\centering
\includegraphics[width=80mm,clip]{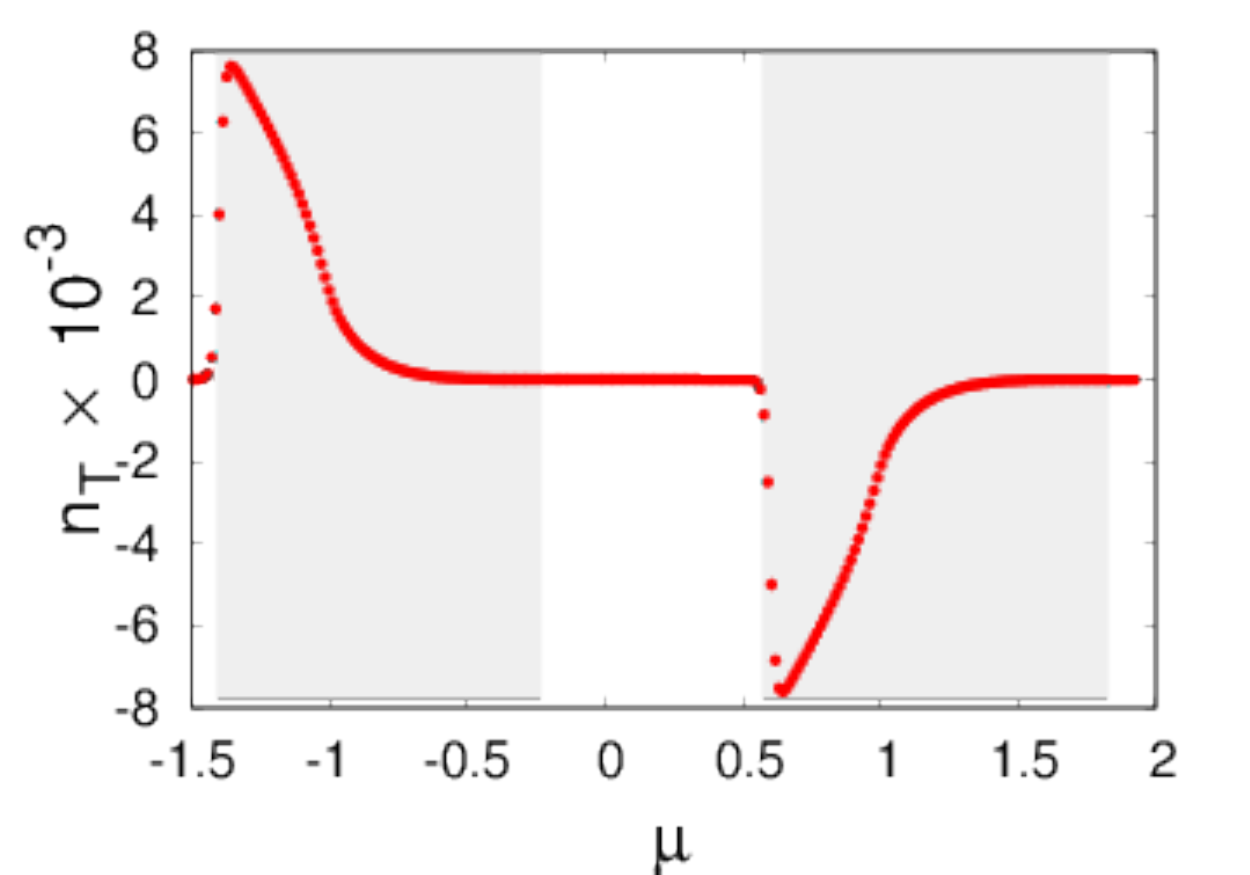}
\caption{(Color online)
The tetrahedral modulation of Fermi surface, $\Braket{n_\mathrm{T}}$. Parameters are $T=0.01$ and $\delta = 0.01$. The shaded area indicates the metallic region. $\Braket{n_\mathrm{T}}$ is finite in the presence of a Fermi surface. 
}
\label{tetrahedral}
\end{figure}

The modulation of Fermi surface may be quantified by the weighted density operator $n_{f}$~\cite{yamase2004},     
\begin{equation}
n_{f}  = \frac{1}{N} \sum_{\bm{k},\alpha} f_{\bm{k}} c_{\bm{k}, \alpha}^{\dagger} c_{\bm{k}, \alpha},
\end{equation}
where $f_{\bm{k}}$ is the weighting function and the index $\alpha$ specifies the internal degree of freedom such as spin and sublattice. For example, $f_{\bm{k}}= \cos{k_x} -\cos{k_y}$ represents the $d_{x^2-y^2}$-wave modulation, and then, the spontaneous ordering of $\Braket{n_{f}}$ is called the $d_{x^2-y^2}$-wave Pomeranchuk instability. 
The tetrahedral modulation of $k_xk_yk_z$ type in BaMn$_2$As$_2$ is given by the following weighted density operator, 
\begin{align}
&n_\mathrm{T}  = \frac{1}{N} \sum_{\bm{k},\tau,\sigma} T_{\bm{k}} c_{\bm{k}, \tau,\sigma}^{\dagger} c_{\bm{k}, \tau,\sigma}, 
\end{align}
where
\begin{align}
&T_{\bm{k}} = \sin{\frac{k_x}{2}}\sin{\frac{k_y}{2}}\sin{\frac{k_z}{2}}.
\end{align}
As we have discussed, the expectation value of $n_\mathrm{T}$ can be regarded as an order parameter of $B_{1u}$ magnetic multipole order in the metallic state. In Figure~\ref{tetrahedral}, we plot $\Braket{n_\mathrm{T}}$ in the AFM state, which is indeed finite in the metallic region.  
The sign of the tetrahedral modulation is naturally opposite between the upper and lower bands because of the sign $\pm$ of the energy spectrum [Eq.~\eqref{energyspectrum}]. 

\begin{figure}[htbp]
\centering
\includegraphics[width=70mm,clip]{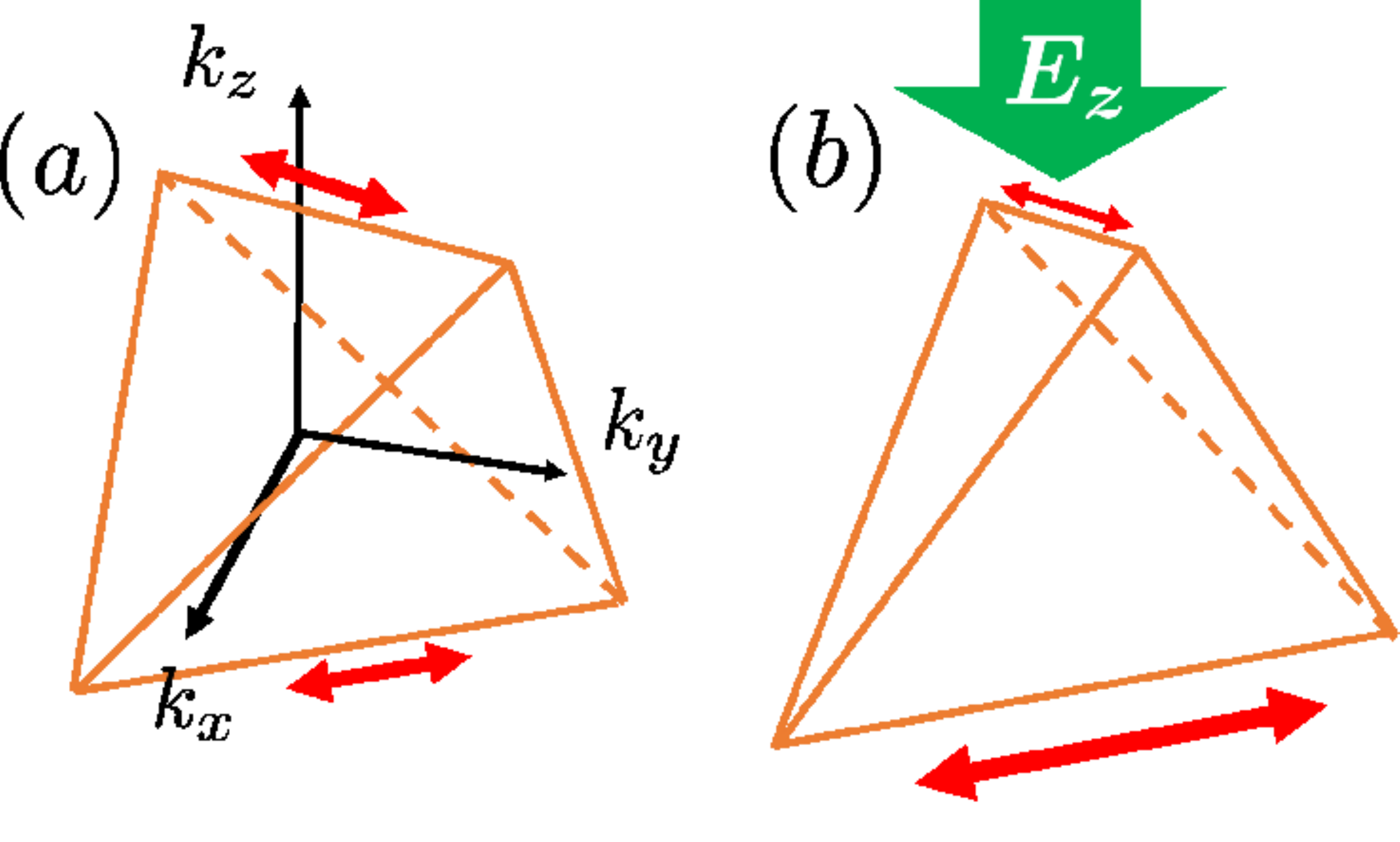}
\caption{(Color online)
Schematic figure for the mechanism of current-induced nematicity. (a) The tetrahedral modulation of a Fermi surface is illustrated. The diagonal nematicity of $k_x k_y$ type is canceled out by the $k_z$-summation. (b) The electric field along the $z$-direction breaks the balance between the $k_z >0$ region and $k_z< 0$ region, giving rise to the $k_xk_y$-diagonal nematic order.}
\label{fermidistort}
\end{figure}

When we look at the Fermi surface on a $k_z =$ constant plane, the diagonal modulation $k_x k_y$ appears. 
However, the $k_x k_y$ modulation is opposite between the $k_z = c$ plane and the $k_z = -c$ plane, and therefore, the summation for $k_z$ results in vanishing in-plane nematic order in the equilibrium state. Now we notice that the applied electric field perpendicular to the $xy$-plane causes the imbalance between $k_z = c$ and $k_z = -c$, which gives rise to the diagonal nematicity in the stationary state. A schematic illustration is shown in Figure~\ref{fermidistort}.   

To investigate the current-induced nematic order, we define a nematic operator $n_\mathrm{D}$ as follows~\cite{comment_nematic}, 
\begin{align}
&n_{\mathrm{D}}  = \frac{1}{N} \sum_{\bm{k},\sigma} D_{\bm{k}} c_{\bm{k}, {\rm A},\sigma}^{\dagger} c_{\bm{k}, {\rm B},\sigma} + h.c.~, 
\label{nematic_operator}
\end{align}
with 
\begin{align}
& D_{\bm{k}} =\sin{\frac{k_x}{2}}\sin{\frac{k_y}{2}}. 
\end{align}
The current-induced nematicity represented by 
\begin{align}
\Braket{n_\mathrm{D}} = \chi_\mathrm{D} E_z, 
\end{align}
is calculated by using Kubo formula, 
\begin{equation}
\chi_{\mathrm{D}}  = \frac{-i e  \hbar}{N} \sum_{\bm{k},\mu.\nu}\frac{\left[	n_\mathrm{D} (\bm{k})	\right]_{\mu \nu}  \left[ v_z (\bm{k}) \right]_{\nu \mu}}{E_{\mu} \left(	\bm{k}	\right)-E_{\nu} \left(	\bm{k}	\right) + i\delta}    \frac{f \left(	E_{\mu} \right)	  -f \left(	E_{\nu}  	\right) }{E_{\mu}  \left(	\bm{k}	\right)-E_{\nu} \left(	\bm{k}	\right)},	  \label{nemele}
\end{equation}
where $n_\mathrm{D} (\bm{k})$ is the band representation of $n_\mathrm{D} $.

\begin{figure}[htbp]
\centering
\includegraphics[width=80mm,clip]{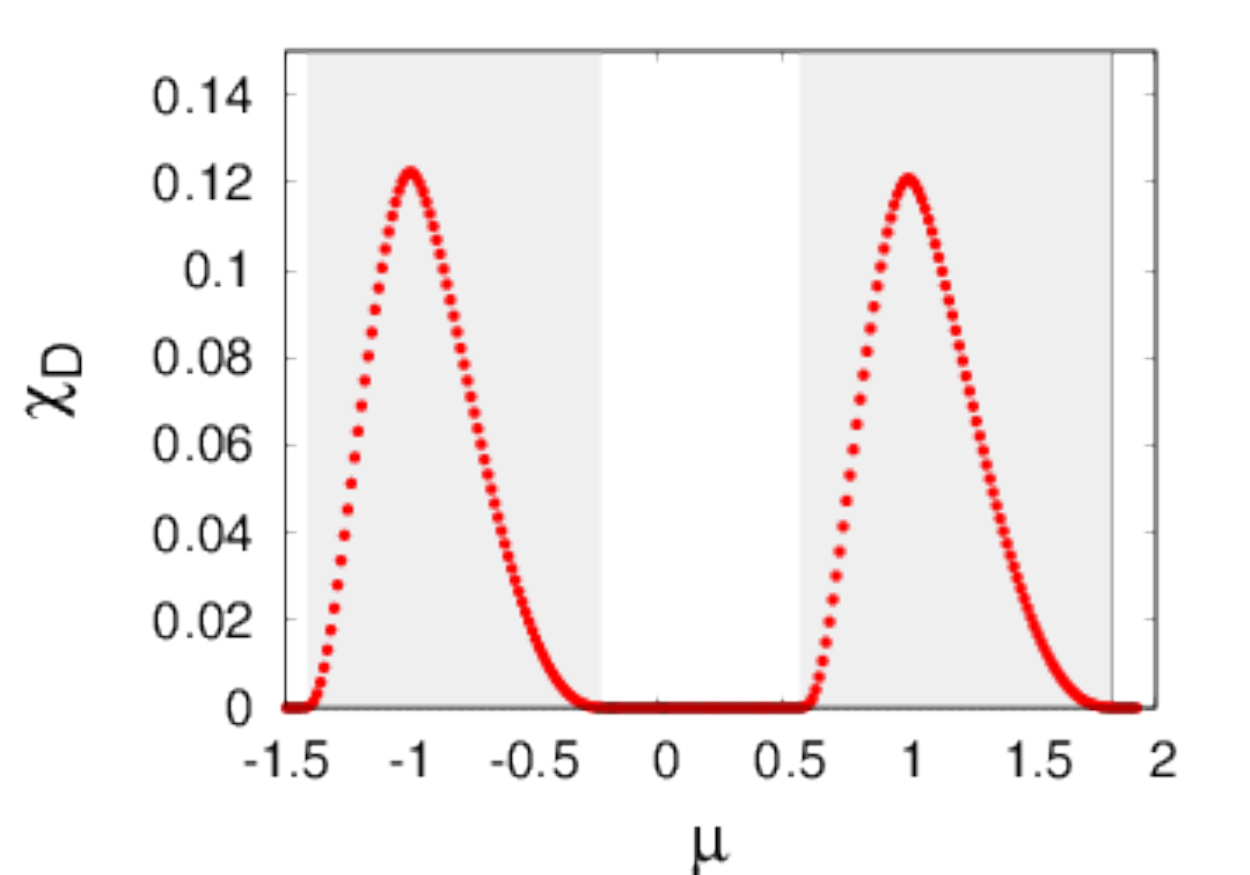}
\caption{(Color online)
The nematic susceptibility to the electric field. Parameters are $T=0.01$ and $\delta = 0.01$, and we take the unit $e\hbar=1$. 
}
\label{nemeleplot}
\end{figure}

Figure~\ref{nemeleplot} shows the numerical result of nematic susceptibility $\chi_\mathrm{D}$. It is indeed shown that the current-induced nematic order occurs in the metallic region. 
Since the nematic operator $n_\mathrm{D}$ is $PT$ even, the nematic susceptibility $\chi_\mathrm{D}$ is determined by intraband contributions as $\alpha^{\rm AF}_{\mu\nu}$ is. Thus, the nematicity is essentially ``current-induced'' and it does not occur in the insulating state. 
This means that the current-induced nematic order is a response characterizing the odd-parity magnetic multipole order in itinerant systems. 

By using the lemma proved in Appendix B, the nematic susceptibility to electric field is obtained as
\begin{equation}
\chi_\mathrm{D}  \simeq \frac{-e \hbar}{\delta V} \sum_{\bm{k},p} \left[ n_\mathrm{D} (\bm{k})\right]_{pp} \frac{ \partial f( E_p ) }{\partial k_z}.
\end{equation}
Thus, $\chi_\mathrm{D} \propto \tau^1$ as expected. 
Note that the nematic susceptibility appears with the same sign between the upper and lower bands, in contrast to the tetrahedral modulation (Figure~\ref{tetrahedral}). This is because the nematic operator is defined by the inter-sublattice hopping. 
By using operators for bonding and anti-bonding orbitals~\cite{harrisonbook},
 	\begin{equation}
	\begin{cases}
	b_{\bm{k}, \sigma} =\frac{1}{\sqrt{2}} \left(	 c_{\bm{k}, A,\sigma} + c_{\bm{k}, B,\sigma}	\right)  & \text{for bonding orbital,}\\
	a_{\bm{k}, \sigma} =\frac{1}{\sqrt{2}} \left(	 c_{\bm{k}, A,\sigma} - c_{\bm{k}, B,\sigma}	\right)  & \text{for anti-bonding orbital,}
	\end{cases}
	\end{equation}
the nematic operator is recast to
\begin{equation}
n_\mathrm{D}= \frac{1}{N} \sum_{\bm{k},\sigma} D_{\bm{k}} \left(	 b_{\bm{k}, \sigma}^{\dagger} b_{\bm{k}, \sigma}- a_{\bm{k}, \sigma}^{\dagger} a_{\bm{k}, \sigma}	\right). 
\end{equation}
Owing to the negative sign in front of $a_{\bm{k}, \sigma}^{\dagger} a_{\bm{k}, \sigma}$, the translation of Fermi surface by electric current induces the nematicity with the same sign in the upper band and the lower band.

\begin{figure}[htbp]
\centering
\includegraphics[width=70mm,clip]{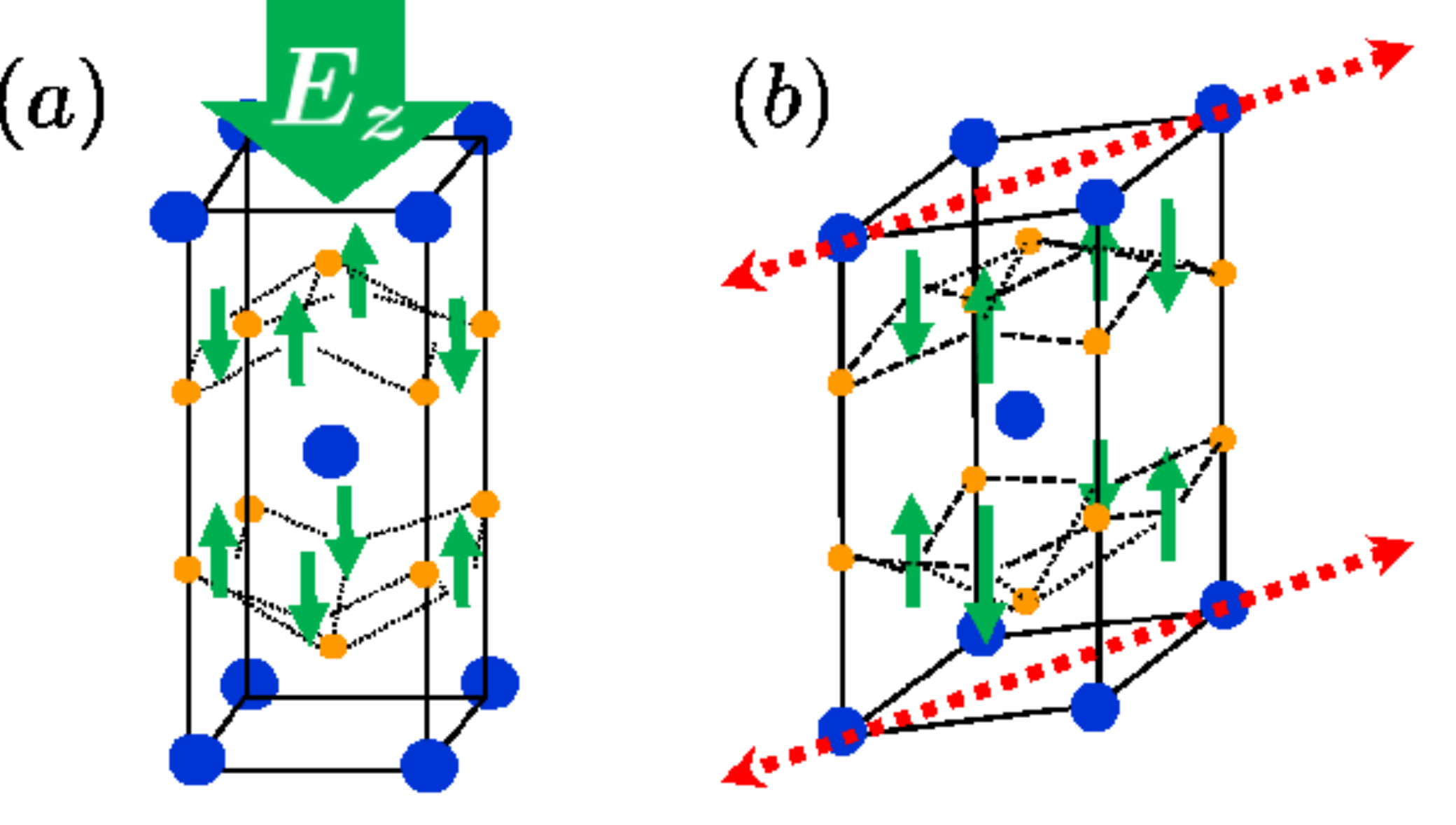}
\caption{(Color online)
Sketch of the current-induced structural transition. The electric field along the \textit{z}-axis induces the electronic nematicity in the \textit{xy}-plane, which leads to the tetragonal-orthorhombic structural transition through electron-lattice couplings.}
\label{nematicdeform}
\end{figure}

Although we have discussed an electronic nematic order so far, the nematicity induces a structural deformation through electron-lattice couplings. Thus, the electric current along the {\it z}-axis induces the lattice structural deformation in the {\it xy}-plane illustrated in Figure~\ref{nematicdeform}. 
The structural nematic order which has been observed in Fe-based 122-compounds~\cite{johnston2010,ishida2009,dai2015} is essentially different from the current-induced nematic order proposed by this work. In $\rm BaFe_2 As_2$, the orthorhombic transition spontaneously occurs at low temperatures. On the other hand, in the odd-parity magnetic multipole state of BaMn$_2$As$_2$, the nematicity is induced by the external electric current. 
Furthermore, the SI symmetry is not broken in the orthorhombic stripe AFM state of BaFe$_2$As$_2$, while the spontaneous SI symmetry breaking plays an essential role in BaMn$_2$As$_2$. As expected from an intuitive explanation for the current-induced nematic order, the electric current along the {\it x}-axis ({\it y}-axis) also induces the structural transition of {\it yz}-type ({\it zx}-type). 

 The structural deformation driven by the electric field can be regarded as a (inverse) piezoelectric effect. 
For insulators, a piezoelectric-coupling constant is given by
\begin{equation}
e_{ijk} = \frac{\partial P_i }{\partial \epsilon_{jk}}\Big|_{\bm{E}=0}, \label{piezo}
\end{equation}
where $\bm{P}$ is an electric dipole moment and $\epsilon_{jk}$ is a strain tensor. Alternatively, it is recast, 
\begin{equation}
s_{ij} = \sum_{k} e_{kij} E_{k},\label{inversepiezo}
\end{equation}
where $s_{ij}$ is a stress tensor and we assume $\epsilon_{ij} =0$. The $D_{2d}$ symmetry allows piezoelectric couplings 
	\begin{equation}
	\begin{split}
		&e_{xyz}=e_{xzy}=e_{yxz}=e_{yzx}, \\
		&e_{zxy}=e_{zyx},
	\end{split}\label{piezo_D2d}
	\end{equation}
and $e_{zxy}$ and $e_{zyx}$ represent the stress in the [110]-direction or the $[\overline{1}10]-$direction induced by the electric field along the [001]-direction. 
This piezoelectric effect is similar to the current-induced nematic order studied in this work. 
However, there are significant differences in their mechanism, symmetry, and manifestation. 
In insulators, the piezoelectric deformation is mainly caused by ionic displacements induced by electric field. 
Then, the polar rank-3 tensor $e_{ijk}$ has the even parity under the TR operation~\cite{restabook}.
On the other hand, the piezoelectricity we propose, namely, the current-induced nematicity, is characteristic of metallic systems, and then the "piezoelectric" tensor $\tilde{e}_{ijk}$ has the odd parity under the TR operation~\cite{commentSHG}. In other words, the direction of the strain, the $[110]$-direction or the $[\overline{1}10]$-direction, is reversed by applying the TR operation. Therefore, the inverse piezoelectric effect is switchable by changing the AFM domain. 
Thus, the metallic magnetic hexadecapole state may be called ``magnetopiezoelectric metal''. 
Interestingly, hole-doped BaMn$_2$As$_2$ realizes such an exotic state which may be useful for device applications.

 We have confirmed that the conventional piezoelectricity does not occur in the magnetopiezoelectric metal from the viewpoint of symmetry. Both of $e_{ijk}$ and $\tilde{e}_{ijk}$ are polar tensors, and require the SI symmetry breaking. The TR even piezoelectric tensor $e_{ijk}$ is forbidden in the magnetic hexadecapole state, since the $PT$ symmetry is preserved. 
Then, the electric field does not directly couple to the strain, but indirectly couples through the electric current. The response is represented by the same form as Eq.~\eqref{inversepiezo}, although the response tensor is replaced by the TR odd one $\tilde{e}_{ijk}$

Recently, a related phenomenon has been proposed by Ref.~\onlinecite{varjas2016}. 
The authors have revealed the electric current generation by a time-dependent strain in metallic systems 
where both of SI symmetry and TR symmetry are broken. 
This is a dynamical and inverse response of the current-induced nematic order which we reveal in this work. 
%

\section{Summary and Discussion}
In this paper, we investigated the odd-parity magnetic multipole order in $\rm BaMn_2As_2$ and clarified characteristic responses. The obtained results are summarized below. 

First, we have classified the magnetic multipole order on the basis of the IRs of point group symmetry, similar to the classification of unconventional superconductivity by Sigrist and Ueda.~\cite{Sigrist-Ueda} The symmetry argument indicates the odd-parity magnetic multipole order in the AFM state of BaMn$_2$As$_2$, which belongs to the $B_{1u}$ IR of $D_{4h}$ point group. Possible multipole moments are magnetic quadrupole moment $M_{22}^+$ and magnetic hexadecapole moment $M_{42}^+$. 

Next, the microscopic analysis of seemingly conventional collinear G-type AFM state in undoped BaMn$_2$As$_2$ and hole-doped Ba$_{1-x}$K$_x$Mn$_2$As$_2$ reveals the leading magnetic hexadecapole order. 
In the hole-doped metallic system, the orbital angular momentum of Mn $3d$ electrons is partially restored, and then the LS-coupling induces the magnetic quadrupole moment $M_{22}^+$ as an admixed odd-parity magnetic order parameter. The microscopic study implies that the local SI symmetry breaking at magnetic sites plays an essential role for the odd-parity magnetic multipole order. Furthermore, we propose a definition of macroscopic order parameter of odd-parity magnetic multipole order, in which ambiguities due to the choice of unit cell are removed. 

Then, we have introduced an effective Hamiltonian and shown electromagnetic responses induced by the odd-parity magnetic multipole order. The ME effect occurs in accordance with the existence of the magnetic quadrupole moment. The AFM Edelstein effect has also been shown, and the electrical switching of magnetic multipole moment has been discussed. Interestingly, the metallic odd-parity magnetic multipole state, where both of TR and SI symmetry are spontaneously broken, shows an asymmetric modulation of Fermi surface. The tetrahedral modulation of $k_xk_yk_z$ type occurs in doped    $\rm BaMn_2As_2$, and induces a counter-intuitive current-induced nematic order. The \textit{in-plane} nematic order is induced by the \textit{out-of-plane} electric current. Thus, the itinerant magnetic hexadecapole state is identified as magnetopiezoelectric metal. These exotic phenomena are derived from the ASOC term arising from local SI symmetry breaking.

Although odd-parity multipole order has been discussed for only a few crystalline materials so far, a variety of magnetic compounds may be identified as odd-parity magnetic multipole state. Indeed, we have revealed that a seemingly conventional AFM state of $\rm BaMn_2As_2$ is identified as the magnetic hexadecapole state. This work is a proposal of magnetic hexadecapole order, although magnetic monopole, toroidal dipole, and magnetic quadrupole compounds have been studied~\cite{astrov1960,spaldin2008,Zhao_Sr2IrO4,Matteo_Sr2IrO4}.
From our analysis of BaMn$_2$As$_2$, we immediately notice that many other compounds show the magnetic hexadecapole order with the magnetic structure similar to BaMn$_2$As$_2$. For instance, we identify other Mn-based 122-systems [BaMn$_2$\textit{Pn}$_2$ (\textit{Pn}=P, Bi)],~\cite{calder2014,brock1994} Cr-based 122-systems [\textit{R}Cr$_2$Si$_2$ (\textit{R}=Ho, Er, Tb) and \textit{Ae}Cr$_2$As$_2$ (\textit{Ae}=Ba, Sr)]~\cite{moze2003,DJsingh2009,filsinger2017,das2017}, Mn-based 112-systems [\textit{X}MnBi$_2$ (\textit{X}=Ca, Sr, Eu)]~\cite{guo2014,masuda2016}, Mn-based 111-systems [KMn\textit{Pn} (\textit{Pn}=As, Sb, Bi)]~\cite{bronger1986,schucht1999}, and Mn-based 1111-systems [LaMnPO and \textit{R}MnAsO (\textit{R}=La, Nd)]~\cite{emery2011,yanagi2009}, as magnetic hexadecapole compounds. 
 The local SI symmetry breaking of magnetic sites and staggered alignment of magnetic moment are satisfactory condition for the odd-parity magnetic multipole order. This condition may be satisfied in various magnetic systems we have not noticed. 
 More elaborate study of odd-parity multipole order will refine understanding of spontaneous parity violation and resulting exotic phenomena in condensed matter.


\begin{acknowledgments}
The authors are grateful to T. Nomoto and S. Sumita for fruitful discussions and comments. 
This work was supported by a Grant-in-Aid for Scientific Research on Innovative Areas ``J-Physics'' (JP15H05884) and ``Topological Materials Science'' (JP16H00991) from the Japan Society for the Promotion of Science (JSPS), and by JSPS KAKENHI Grants (Numbers
JP15K05164 and JP15H05745).
\end{acknowledgments} 

\appendix
\section{Derivation of the effective model}
Five-orbital tight-binding Hamiltonian for Mn $3d$ orbitals is represented by
	\begin{equation}
	\mathcal{H} = \mathcal{H}_{\rm even} + \mathcal{H}_{\rm odd} + \mathcal{H}_{\rm LS} + \mathcal{H}_{\rm CEF} + \mathcal{H}_{\rm AFM}, \label{fullmodel}
	\end{equation}
where $ \mathcal{H}_{\rm even}$ and $\mathcal{H}_{\rm odd}$ are hopping terms with even- and odd-parity under the SI operation, respectively. The LS-coupling term is written as
 \begin{equation}
\mathcal{H}_{\rm LS} = \lambda \sum_{i,\tau }\bm{l}_{i,\tau}\cdot \bm{s}_{i,\tau},
\end{equation}
where $\lambda$ is the coupling strength, $\bm{l}$ ($\bm{s}$) is orbital (spin) angular momentum operator, and the label $i$ and $\tau$ indicate the site and sublattice index, respectively. The crystalline electric field term $\mathcal{H}_{\rm CEF}$ which mainly arises from the ligand field due to As atoms (Figure~\ref{cluster}) gives rise to the level splitting of Mn $d$ orbitals. The $d$ levels are classified by the local point group $D_{2d}$ of Mn sites,
\begin{equation}
\underbrace{d_{z^2}}_{A_1} + \underbrace{d_{xy}}_{B_1}+\underbrace{d_{x^2-y^2}}_{B_2}+\underbrace{d_{yz},d_{zx}}_{E},  
\end{equation}
where the IR of the point group is indicated for each $d$ level. We here neglect electron correlation effects in the AFM state and take into account the molecular field term, 
\begin{equation}
\mathcal{H}_{\rm AFM} = \sum_{i,\tau} -2 h \tau_z s^z_{i,\tau}, 	
\end{equation}
where $\bm{\tau}$ is the Pauli matrix acting on the sublattice space.

Now we derive the single band Hamiltonian for the valence band. Because the LS-coupling is small in $3d$ electron systems, the Russell-Saunders picture is appropriate. The crystalline electric field is much larger than the LS-coupling. Therefore, we can perturbatively treat the LS-coupling term. Then, the eigenstate of atomic Hamiltonian $\mathcal{H}_{\rm CEF} +\mathcal{H}_{\rm LS} + \mathcal{H}_{\rm AFM}$ for mainly $d_{x^2-y^2}$ orbital is obtained as
\begin{align}
		\ket{\sigma_z=\pm ,\tau_z } =& \ket{d_{x^2-y^2}, \sigma_z,\tau_z} + \frac{i\lambda \sigma_z}{\Delta_1}\ket{d_{xy}, \sigma_z ,\tau_z} \notag \\
		 &- \frac{i\lambda}{2 \left(  \Delta_2- 2h \sigma_z \tau_z \right)	}\ket{d_{yz}, -\sigma_z,\tau_z} \notag \\
		 &+ \frac{\lambda \sigma_z}{2\left(	\Delta_2 -2h \sigma_z \tau_z	\right) }\ket{d_{zx},-\sigma_z,\tau_z},
\end{align}
where $\bm{s} =\frac{1}{2}\bm{\sigma}$, and $\Delta_1$ $(\Delta_2)$ is the energy level of $d_{xy}$ orbital ($d_{yz}$ and  $d_{zx}$ orbitals) from the level of $d_{x^2-y^2}$ orbital.

Projecting the five-orbital model [Eq.~\eqref{fullmodel}] to the Hilbert space spanned by $\ket{\sigma_z ,\tau_z }$, we obtain the projected Hamiltonian as Eq.~\eqref{efhamiltionian0}. The coupling constants of the ASOC term and the SSOC term for the A sublattice, $\bm{g}_{A}\cdot \bm{\sigma}$, are obtained as
\begin{align}
&\alpha_{1}= \frac{2t_{zx,1}\lambda \Delta_2 }{\Delta_2^2-4h^2}, \label{a1} \\
&\alpha_{2}= \frac{4\left( t_{zx,2}^{(1)}-t_{zx,2}^{(2)}\right) \lambda \Delta_2}{\Delta_2^2 -4h^2}, \label{a2} \\
&\alpha_{3}= \frac{16t_{xy,3}\lambda}{\Delta_1},\label{a3} \\
&\beta= -\frac{8\left( t_{zx,2}^{(1)}+t_{zx,2}^{(2)}\right) \lambda h}{\Delta_2^2 -4h^2}. \label{b}
\end{align}
The hopping integrals are written as
\begin{align}
&-t_{zx,1} = \bra{d_{zx} ,(a,0,0)}\mathcal{H}_{\rm kin}\ket{d_{x^2-y^2},(0,0,0)}, \label{t1} \\
&-t_{zx,2}^{(1)} = \bra{d_{zx} ,(\frac{a}{2},\frac{a}{2},\frac{c}{2})}\mathcal{H}_{\rm kin}\ket{d_{x^2-y^2},(0,0,0)}, \label{t2a}\\
&-t_{zx,2}^{(2)} = \bra{d_{zx},(\frac{-a}{2},\frac{-a}{2},\frac{-c}{2})}\mathcal{H}_{\rm kin}\ket{d_{x^2-y^2},(0,0,0)},\label{t2b} \\
&-t_{xy,3} = \bra{d_{xy},(\frac{a}{2},\frac{a}{2},\frac{c}{2})}\mathcal{H}_{\rm kin}\ket{d_{x^2-y^2} ,(0,0,0)}, \label{t3}
\end{align}
where $\mathcal{H}_{\rm kin} = \mathcal{H}_{\rm even} + \mathcal{H}_{\rm odd}$, $\ket{d_\gamma , (x,y,z)}$ denotes the orbital wave function of Mn $3d_\gamma$ orbital on the A sublattice, and $(x, y, z)$ is the Cartesian coordinates.

The entanglement of spin and orbital due to the LS-coupling results in the $\bm{k}$-dependent spin-orbit coupling terms. According to Eqs.~\eqref{a1}-\eqref{b}, the in-plane components of the $g$-vector originate from the hybridization of $\ket{d_{x^2-y^2},\sigma_z}$ with $\ket{d_{zx} (d_{yz}),-\sigma_z}$. Therefore, the coupling constants specifying the in-plane component, namely, $\alpha_1$, $\alpha_2$ and $\beta$, are suppressed by the large AFM molecular field. On the other hand, the out-of-plane component of the ASOC term, $g'_z(\bm{k}) \sigma_z$, is robust against the AFM order. 

The SSOC term, $\bm{g}''(\bm{k}) \cdot {\bm \sigma}$, is induced by the AFM order, although it disappears in the paramagnetic state.  
This term does not play an essential role for the electromagnetic responses studied in this paper. 
Because the SSOC term remains finite in the reference state shown in Figure~\ref{virtualcrystal}(a), it does not induce the ME effect characteristic of the odd-parity magnetic multipole state. 
From  Eq.~\eqref{energyspectrum}, we notice that the coupling between the ASOC term and the SSOC term also induces the tetrahedral modulation of the band structure, which is given by
\begin{align}
& {\bm g'}({\bm k}) \cdot {\bm g''}({\bm k}) = \notag \\
& 2\alpha_1 \beta \left( \sin{k_y} \sin{\frac{k_x}{2}} \cos{\frac{k_y}{2}}-  \sin{k_x} \sin{\frac{k_y}{2}} \cos{\frac{k_x}{2}} \right) \sin{\frac{k_z}{2}}.
\label{crossterm}
\end{align}
Although the current-induced nematicity studied in Sec.~VI also arises from this term, it is a higher order correction with respect to the LS-coupling constant $\lambda$. Since the LS-coupling is much smaller than the AFM molecular field and the crystalline electric field, Eq.~\eqref{crossterm} is negligible compared with the leading order term proportional to $h \alpha_3$. Therefore, we neglect the SSOC term in Secs.~V and VI.

\section{Lemma for Kubo formula}
Electromagnetic responses in the linear response region are generally given by Kubo formula. When the Hamiltonian is represented by a quadratic form of one-body operators, the response function for uniform and static perturbation is obtained as a simple form, 
\begin{equation}
\chi_{AB } = C \sum_{\bm{k}, p,q} \frac{ \left[ A (\bm{k})\right]_{pq} \left[ B (\bm{k})\right]_{qp} }{E_{p} (\bm{k})-E_{q} (\bm{k}) +i\delta} \frac{f( E_{p})  -f( E_{q} )  }{ E_{p} (\bm{k})-E_{q} (\bm{k}) }, \label{response}
\end{equation}
where $C$ is a constant factor, $p$ and $q$ are band indices, $ \left[ A (\bm{k}) \right]_{pq}$ and $\left[ B (\bm{k}) \right]_{qp}$ are band representation of uniform operators $A$ and $B$, respectively. The band energy is denoted by $E_{p}({\bm k})$, $\delta$ is a constant scattering rate, and $f(E)$ is the Fermi distribution function. For instance, the electric conductivity tensor $\sigma_{\mu\nu}$ is obtained by assigning the current operators $j_\mu$ and $j_\nu$ to $A$ and $B$, respectively. 

Here, we consider the Hamiltonian which preserves the $PT$ symmetry. The $PT$ symmetric system has at least double degeneracy at each momentum $\bm{k}$ (Kramers pair), and single particle states are labeled by $\sigma_z =\pm$ with Pauli matrix acting on the degenerate Hilbert space. The Bloch states of the Kramers pair are transformed to each other by the $PT$ operation,
	\begin{equation}
	PT \ket{n,\bm{k},\sigma} = \left(	i\sigma_2	\right)_{\sigma' \sigma}  \ket{n, \bm{k}  ,\sigma'},
	\end{equation}
where $\ket{n,\bm{k},\sigma}$ is denoted by the band index $n$, crystal momentum $\bm{k}$, and (pseudo-)spin $\sigma$. The band index $p$ (and $q$) is specified by the combination of $n$ and $\sigma$ ($n'$ and $\sigma'$). The intraband contributions to the response function Eq.~\eqref{response} come from pairs $(p,q)$ with $n=n'$. Thus, intraband contributions are regarded as intra-Kramers pair contributions. On the other hand, interband contributions are given by bands with nonequivalent energy, $n \ne n'$.


Now we show a lemma about the relation between the $PT$ parity of $A$ and  $B$ and the response function: 
\begin{itemize}
\item
When the product of $PT$ parity of operators $A$ and $B$ is odd, the response function $\chi_{AB}$ is determined by the interband contributions. 
\item
When the product of $PT$ parity is even, the response function is given by the intraband contributions. 
\end{itemize}
To prove the lemma, we consider two Kramers pairs protected by the $PT$ symmetry, $\ket{n,\bm{k},\sigma}$, $\ket{m,\bm{k},\sigma}$. By inserting the $PT$ operator the matrix element $\left[ A (\bm{k})\right]_{pq}$ for $(p,q) =\left[ ( n  ,\sigma ),  (  m  ,\sigma' ) \right]$ is transformed as 
	\begin{align}
		&\left[ A (\bm{k}) \right]_{(n,\sigma),(m,\sigma')} \\
		& =\left[   \tilde{A} (\bm{k}) \right]_{(m,\sigma''),(n,\sigma''')} \left(	i\sigma_2	\right)^\dagger_{\sigma' \sigma''}\left(	i\sigma_2	\right)_{\sigma''' \sigma}\\
	& =\left[   A (\bm{k}) \right]_{(m,\sigma''),(n,\sigma''')} \left(	i\sigma_2	\right)^\dagger_{\sigma' \sigma''}\left(	i\sigma_2	\right)_{\sigma''' \sigma} (-1)^{P_A}, \label{reciprocal}
	\end{align}
where $\tilde{A}=(PT) A (PT)^{-1}$ and $P_A$ denotes the $PT$ parity of the Hermitian operator $A$. Here, we use $(PT) \bm{k} =\bm{k}$ and the anti-Hermitian property of $PT$-operator
	\begin{eqnarray}
	&\bra{\phi} A\ket{\psi} = \bra{\tilde{\psi}}(PT) A^\dagger (PT)^{-1} \ket{\tilde{\phi}}, \\
	& \ket{\tilde{\phi}}=PT\ket{\phi}, \ket{\tilde{\psi}}=PT\ket{\psi}.
	\end{eqnarray}
By using the relation Eq.~\eqref{reciprocal}, the summation for the band index in Eq.~\eqref{response} simplified. 

First, the intra-Kramers pair contributions [$ p,q  = ( n  ,\sigma ),  (  n  ,\sigma' )$] to the response function are given by 
\begin{align}
&\chi_{AB}^{({\rm intra})} = \notag \\
& \sum_{n,  \bm{k}  } \frac{C}{i\delta} \frac{\partial f(E) }{\partial E}\Big|_{E_{n}({\bm k}) } \sum_{\sigma,\sigma'}  \left[ A (\bm{k})\right]_{(n,\sigma),(n,\sigma')}\left[ B (\bm{k})\right]_{(n,\sigma'),(n,\sigma)}. 
\end{align}
The r.h.s is transformed by 
\begin{align}
&\sum_{\sigma,\sigma'}  \left[ A (\bm{k})\right]_{(n,\sigma),(n,\sigma')}\left[ B (\bm{k})\right]_{(n,\sigma'),(n,\sigma)} \\
&= \sum_{\sigma,\sigma'}  \left[ A (\bm{k})\right]_{(n,\sigma'),(n,\sigma)}\left[ B (\bm{k})\right]_{(n,\sigma),(n,\sigma')}  (-1)^{P_{AB}},
\end{align}
where $P_{AB} =P_A+P_B$. Hence, we obtain 
\begin{equation}
\chi_{AB}^{({\rm intra})} =(-1)^{P_{AB}} \chi_{AB}^{({\rm intra})}. \label{intraKramers}
\end{equation}

Similarly, the inter-Kramers pair contributions [$ p,q  = (n,\sigma), (m,\sigma')$ with $n\neq m$] are simplified as follows. We divide the inter-Kramers pair contributions 
	 \begin{equation}
	\chi_{AB}^{({\rm inter})} =\chi_{AB}^{({\rm inter,odd})} + \chi_{AB}^{({\rm inter,even})},  
	\end{equation}
with  
	\begin{align}
	\chi_{AB}^{({\rm inter,odd})}&= \sum_{n,m,  \bm{k}  } \alpha \frac{f(E_n) -f(E_m)}{\Delta_{nm}} \frac{\Delta_{nm} }{\Delta_{nm}^2 + \delta^2  } \notag \\ 
		&\times \sum_{\sigma,\sigma'}   \left[ A (\bm{k})\right]_{(n,\sigma),(m,\sigma')}\left[ B (\bm{k})\right]_{(m,\sigma'),(n,\sigma)}, \\
	\chi_{AB}^{({\rm inter,even})}&= \sum_{n,m,  \bm{k}  } \alpha \frac{f(E_n) -f(E_m)}{\Delta_{nm}} \frac{-i\delta }{\Delta_{nm}^2 + \delta^2  } \notag \\
	& \times  \sum_{\sigma,\sigma'}   \left[ A (\bm{k})\right]_{(n,\sigma),(m,\sigma')}\left[ B (\bm{k})\right]_{(m,\sigma'),(n,\sigma)}, 
	\end{align}
and $\Delta_{nm} \equiv E_n (\bm{k})-E_m (\bm{k})$. Then, we obtain
	\begin{align}
	\chi_{AB}^{({\rm inter,odd})} =-(-1)^{P_{AB}} \chi_{AB}^{({\rm inter,odd})},  \label{interodd}\\
	\chi_{AB}^{({\rm inter,even})} =(-1)^{P_{AB}} \chi_{AB}^{({\rm inter,even})}. \label{intereven} 
	\end{align}
An arbitrary scattering rate $\delta$ is set to zero for inter-band contributions, as usual. Thus, we have 
\begin{equation}
\chi_{AB}^{({\rm inter})} = - (-1)^{P_{AB}} \chi_{AB}^{({\rm inter})}. \label{interKramers}
\end{equation}

Eqs.~\eqref{intraKramers} and \eqref{interKramers} clarify the relation between the combined $PT$ parity of $A$ and $B$ and the response function. 
\begin{align}
& \chi_{AB}^{({\rm intra})} =0 \hspace{5mm} \text{for odd $P_{AB}$}, \\
& \chi_{AB}^{({\rm inter})} =0 \hspace{5mm} \text{for even $P_{AB}$}. 
\end{align}
Thus, the response function is given by the intraband contributions for even $P_{AB}$, while it is given by the interband contributions for odd $P_{AB}$. This is the lemma which is proved in this section. Our results in Secs. V and VI have been discussed on the basis of the lemma. 

The scaling with respect to the scattering rate is also obtained from the proof. 
\begin{align}
& \chi_{AB}  \propto \delta^{-1} \hspace{5mm} \text{for even $P_{AB}$}, \\
& \chi_{AB}  \propto 1 \hspace{8.8mm} \text{for odd $P_{AB}$}. 
\end{align}
The response function $\chi_{AB}$ for even $P_{AB}$ originates from the deformation of Fermi surface, and therefore, disappears in the insulating state lacking Fermi surface. On the other hand, for odd $P_{AB}$, the response function comes from the deformation of wave function. Then, a finite response function may be obtained even in the insulating state. 

The lemma is an extension of Onsager's reciprocity relation. It is straightforward to prove other relations for response functions, for instance ensured by TR symmetry.

\section{Nematic operators}
In this paper we adopt the nematic operator $n_\mathrm{D}$ in Eq.~\eqref{nematic_operator}. However, the nematic operator quantifying the deformation of Fermi surface is not unique. Indeed, we may consider another nematic operators 
\begin{eqnarray}
&n_\mathrm{D1}  =\dfrac{1}{N} \sum_{\bm{k},\tau,\sigma} 	D_{\bm{k}}' c_{\bm{k}, \tau,\sigma}^{\dagger} c_{\bm{k}, \tau, \sigma}, \\
&n_\mathrm{D2} = \dfrac{1}{N}\sum_{\bm{k},\tau,\sigma} 	D_{\bm{k}}'' c_{\bm{k}, \tau,\sigma}^{\dagger} c_{\bm{k}, \tau, \sigma},
\end{eqnarray}
with weighting functions $D_{\bm{k}}' = \sin{k_x} \sin{k_y}$ and $D_{\bm{k}}'' = \sin{(k_x/2)} \sin{(k_y/2)} \cos{(k_z/2)}$. These nematic operators correspond to the next-nearest-neighbor intralayer hopping and nearest-neighbor interlayer hopping, respectively. The nematic order parameters $\Braket{n_\mathrm{D1}}$ and $\Braket{n_\mathrm{D2}}$ belong to the same IR ($B_{2g}$) of the $D_{4h}$ point group as $\Braket{n_\mathrm{D}}$. Hence we do not have any unique definition for the nematic operator. In this paper we adopt $n_\mathrm{D}$ corresponding to the nearest-neighbor Mn-Mn hopping. In fact, the nematic susceptibility defined for the nematic operator $n_\mathrm{D}$ shows larger value than those for $n_\mathrm{D1}$ and $n_\mathrm{D2}$.


\bibliography{reference}

\begin{thebibliography}{33}
\bibitem{comment1}
A local orbital of Mn atoms with mixed parity can be formed by hybridization of Mn 3$d$ orbitals with Mn $p$ or $f$ orbitals. However, the energy level of Mn $p$ and $f$ orbitals are far from the Fermi level. Hence, we neglect the parity mixing on the Mn atoms.
\bibitem{comment_nematic}
Definition of nematic operator is not unique. We may adopt another nematic operators as shown in Appendix C. The current-induced nematicity occurs irrespective of the choice of nematic operator.
\end{thebibliography}

\clearpage
\end{document}